\documentclass[fleqn,usenatbib]{mnras}

\usepackage[T1]{fontenc}
\usepackage{ae,aecompl}

\usepackage{graphicx}	
\usepackage{amsmath}	
\DeclareMathOperator{\sech}{sech}

\usepackage{amssymb}	

\usepackage{txfonts} 

\usepackage{epstopdf}
\usepackage{journals}

\usepackage[none]{hyphenat}
\graphicspath{{images_lr/}{images/}}
\DeclareGraphicsExtensions{.pdf,.png,.jpg, .jpeg}
\usepackage{natbib}
\usepackage{color}
\usepackage{caption}

\usepackage{threeparttable}

\title[What determines the flatness of X-shaped structures in edge-on galaxies]{What determines the flatness of X-shaped structures in edge-on galaxies}
\author[Anton A. Smirnov and Natalia Ya. Sotnikova]
{Anton A. Smirnov$^{1,2}$\thanks{E-mail:
zeleniikot@gmail.com (AAS)} and Natalia Ya. Sotnikova$^{1}$\\
$^{1}$St. Petersburg State University,
Universitetskij pr.~28, 198504 St. Petersburg, Stary Peterhof, Russia\\
$^{2}$Central (Pulkovo) Astronomical Observatory of RAS, Pulkovskoye Chaussee 65/1, 196140 St. Petersburg, Russia\\
}

\date{Accepted XXX. Received YYY; in original form ZZZ}

\pubyear{2017}

\begin{document}
\label{firstpage}
\pagerange{\pageref{firstpage}--\pageref{lastpage}}
\maketitle

\begin{abstract}
Recent observational studies of X-shaped structures revealed that values of their opening angles lie in a narrow range: from 20$^\circ$ to 43$^\circ$ with smaller X-shaped structures appearing to follow a characteristic opening angle $\sim$ 27$^\circ$ -- 31$^\circ$. We use self-consistent numerical simulations to uncover what parameters of host galaxies govern the opening angle spread. We constructed a series of equilibrium models of galaxies with high spatial resolution, varying the dark halo and bulge contribution in the overall gravitational potential, the initial disc thickness of models and the Toomre parameter $Q$ and followed their evolution for almost 8~Gyrs. Each model demonstrated the formation of clear X-structures with different flatness. We have obtained that opening angles lie in the range from 25$^\circ$ to 42$^\circ$ throughout the entire evolution. These values are roughly consistent with observational data. The greatest variation in the opening angles is obtained by varying the mass of the dark halo. The initial thickness of the disc and the Toomre parameter are responsible for smaller variations of the angle and shorter X-structures. An increase of both parameters changes the morphology of in-plane bars and X-structures. In some cases we observed even double X-structures. The main effect of the bulge is to prevent buckling at intermediate and late stages of the disc evolution. Comparison of models with different halo masses indicates that the smallest observable values of opening angles can be associated with the presence of a heavy dark halo (more than 3 masses of the disc within the optical radius). 
\end{abstract}


\section{Introduction}
X-structures are clearly visible on images of some edge-on galaxies. For example, on the SDSS image of NGC~4469, one can see four bright rays coming from the centre of the stellar disc. These rays stand out well against the galaxy image. They are almost symmetric with respect to a mid-plane of the stellar disc and a plane passing through the centre of the disc perpendicular to the mid-plane of the stellar disc. The whole structure resembles a cross that strikes out the image of the host galaxy seen edge-on.
\par
The cross is the brightest feature of another structure --- a boxy/peanut-shaped (B/PS) bulge (see, for example, \citealt{Aronica_etal2003}). \cite{Combes_Sanders1981} showed via numerical simulations that a box shape morphology of a galaxy can be associated with a bar that has grown in the vertical direction. Further numerical studies confirmed that after the bar forms it tends to thicken quickly \citep{Combes_etal1990}. After a period of rapid growth and buckling in the vertical direction the bar seen edge-on has a boxy/peanut shape \citep{Combes_etal1990,Pfenniger_Friedli1991,Raha_etal1991,Athanassoula_Misiriotis2002,Oneil_Dubinski2003,Athanassoula2005,MartinezValpuesta_etal2006,Debattista_etal2006}. In models with high resolution an X-structure is distinguished as two diagonals of the boxy structure in the grayscale pictures (see, for example, \citealt{Salo_Laurikainen2017_v2}).
\par
Two possible scenarios of formation of B/PS structures in disc galaxies were proposed. \cite{Raha_etal1991} suggested that such a structure can occur due to violent buckling instability of a bar similar to the global bending instability of a fire-horse type of a stellar disc \citep{Toomre1966,Poliachenko_Shukhman1977,Araki1985,Merritt_Sellwood1994}. 
\textcolor{black}{
Bending instability arises in systems with flattened velocity ellipsoid when the vertical-to-radial velocity dispersion ratio $\sigma_z/\sigma_R$ drops below some critical threshold \citep{Poliachenko_Shukhman1977,Araki1985}. In this case a star travels one wavelength of the bending perturbation faster than it manages to make one oscillation in the vertical direction, which leads to an increase in the perturbation by collective processes. The subsequent thickening of the disc results in saturation of the instability.
}
Another explanation is a gradual vertically symmetric growth of a bar via resonant trapping of individual stars \citep{Quillen2002,Quillen_etal2014}. In this case the incidence of B/PS structures in models containing a vertical 2:1 resonance is interpreted in terms of destabilization of in-plane orbits through the bifurcation of orbit families at the resonance (\citealt{Combes_etal1990},\mbox{
\citealt{Pfenniger_Friedli1991}}). A growing bar is populated by periodic orbits with typical vertical frequencies which commensurate with pattern speed of a bar. Detail studies of the morphology of orbits in the rotating (rigid or frozen) gravitational potential revealed orbits which could constitute the ``backbone'' of the B/PS structures \citep{Combes_etal1990,Pfenniger_Friedli1991,Patsis_etal2002,Patsis_Katsanikas2014,Portail_etal2015}, thus reaffirming the concept of resonant trapping. Nevertheless, a thin bar does exhibit the bend at least at the early stage of its evolution \citep{Raha_etal1991} as has been proved by the analysis of bending modes \citep{Sotnikova_Rodionov2003}. Moreover, \cite{Friedli_Pfenniger1990} performed a series of $N$-body simulations and found a slight symmetry breaking in $z$ direction to be necessary for creating the boxy/peanut shapes. When they artificially suppressed the symmetry in $z$ direction, the formation of the box shape was significantly slowed down or even stopped \citep{Pfenniger_Friedli1991}. The rapid break of the vertical symmetry in a growing bar has been observed in many simulations not only at early stages of the bar evolution \citep{Berentzen_etal1998,MartinezValpuesta_etal2004,Debattista_etal2006} but at intermediate and even late stages, the so-called recurrent buckling \citep{Oneil_Dubinski2003,MartinezValpuesta_etal2006,Saha_etal2013}. At first glance, it evidences against resonant trapping. Moreover, direct detection of ongoing buckling in a few barred spirals visible at intermediate inclination has been recently made by \citet{Erwin_Debattista2016,Erwin_Debattista2017} and \citet{Li_etal2017}. But the statistics are so poor that the authors do not rule out the possibility that some B/PS bulges could be a result of alternate, symmetric growth mechanisms like resonant trapping \citep{Erwin_Debattista2016,Erwin_Debattista2017}.
\par
As for the X-structure, \cite{Friedli_Pfenniger1990} and \cite{Pfenniger_Friedli1991} believed that it is an optical illusion in grayscale photos because this shape does not correspond to the isophotes. The illusion is due to the fact that the eyes tend to perceive the image gradients rather than absolute values. Indeed most of the original images of B/PS bulges show only a boxy shape but X-structures are clearly seen in the unsharp-masked images \citep{Aronica_etal2003,Bureau_etal2006,Laurikainen_etal2014}. \cite{Patsis_etal2002} found in several simple models the families of periodic orbits that can maintain the X-shape.
\par
Whatever is the formation mechanism of B/PS bulges, there are no strong differences in the resulting end-stage structures. This conclusion can be particularly applied to X-structures, which look very symmetrical in edge-on galaxies \citep{Ciambur_Graham2016,Savchenko_etal2017}.
From an observational point of view, X-structures are characterised by two parameters: an opening angle (an angle between the ray and the major axis of the galaxy) and the length of rays. The tangent of the opening angle gives the flatness of the X-structure. New observational data indicate that parameters of X-structures for real galaxies lie within a narrow range. \cite{Ciambur_Graham2016} studied X-structures in 11 edge-on galaxies and found that the opening angle falls within
$20^{\circ}$ and $43^{\circ}$, except one galaxy from the sample that has a very small opening angle (15$^\circ$) and a very flattened morphology.
\citet{Savchenko_etal2017} give  approximately the same range from $20^{\circ} \pm 2^{\circ}$ to $38^{\circ} \pm 2^{\circ}$ for their sample of 22 edge-on galaxies.
\cite{Laurikainen_Salo2017} thoroughly investigated two large samples of X-shaped and barlens galaxies, which are face-on counterparts of the first objects. They found X-structures even in galaxies at intermediate inclinations and measured their flatness. If we translate the data by  \cite{Laurikainen_Salo2017} into opening angles, then for galaxies close to the edge-on position ($i>70^{\circ}$), the angle spread will be in the range from $24^{\circ}$ to $45^{\circ}$ with a small fraction of galaxies with larger angles. \cite{Savchenko_etal2017} showed that the distribution over the opening angles is contaminated by the projection effect when a bar is viewed from different position angles. However, the most flattened structures appear to be observed when the bar is located side-on.
\par
In this work, we try to understand what determines the observed range of opening angle values besides the projection effects and to find physical reasons which lead to an increase or a decrease in the value of the opening angle. We focused on following parameters: the relative mass of a dark halo, the mass concentration (presence of a bulge), an initial disc thickness and a Toomre parameter $Q$. The main reasons for this choice are the following. 
\par
It was proved by numerical studies that the evolution and final structure of stellar discs, which are slightly unstable against in-plane and bending perturbations, is determined to a great extent by the relative contribution of the spherical component (mainly a dark halo) to the overall gravitational field within the stellar disc. For example, \citet{Athanassoula_Misiriotis2002} found that in models with the high relative halo mass, a bar is stronger than in models with a light halo and it shows a more pronounced X-shape. 
The final disc thickness also depends on the halo mass. In the absence of separate agents causing out-plane scattering of stars (GMCs, nearby external galaxies), the outer regions of stellar discs in isolated galaxies can be heated by the bending instability \citep{Rodionov_Sotnikova2013}. It heats the disc up to the level corresponding to the linear criterion of the bending instability --- $\sigma_z/\sigma_R>0.3-0.4$ \citep{Poliachenko_Shukhman1977,Araki1985}. If a flat system is stable against in-plane and out-plane perturbations its minimal thickness $z_\mathrm{d}$ will be determined by the relative mass of a dark halo \citep{Zasov_etal1991,Zasov_etal2002}. The existence of the thinnest discs is possible only if there are very massive dark halos \citep{Sotnikova_Rodionov2006,Rodionov_Sotnikova2013}. Extremely massive dark halos can even completely suppress buckling \citep{Saha_etal2013}. The presence of a massive spheroidal bulge alters the morphology of a bar and an X-structure \citep{Salo_Laurikainen2017_v2}. Moreover, buckling can be substantially damped in the presence of a compact bulge \citep{Sotnikova_Rodionov2005}. The same effect is observed if the disc is already vertically hot. \citet{Fragkoudi_etal2017} give an example of X-structures for initially thin and thick discs. The bar in a thick disc is shorter, thicker and the final X-structure is less pronounced (see also \citealt{Debattista_etal2005,Li_Shen2015}).
\par
Either the leading mechanism of the X-structure formation is the resonant trapping or the fire-horse instability, the mass of the dark matter determines the velocity distribution of stars, i.e. characteristic frequencies of stellar orbits and relation between velocity dispersions, and thereby determines the shape of an X-structure. By the same reasoning, other parameters such as an initial disc thickness, a Toomre parameter and the contribution of a bulge to the overall gravitation field also may contribute to the morphology of an X-structure.
\par
To analyse the dependence of an  X-structure opening angle on the chosen parameters, we conducted a series of numerical simulations. We constructed a set of equilibrium $N$-body multi-component models with different values of the relative mass of the dark matter, mass of a bulge, initial disc thickness and $Q$, and solved the equations of motions for a period of time comparable to the lifetime of real galaxies. We assume that the major axis of the bar is perpendicular to the line of sight and traced the evolution of X-structure parameters with time. 
\par
Before comparing X-structures in different models, we analysed the numerical effects associated with relaxation in the vertical direction of modelled galaxies. The relaxation in the vertical direction manifests itself in thickening of a stellar disc and an increase of the vertical-to radial velocity dispersion ratio \citep{Sellwood2013,Rodionov_Sotnikova2013}. For example, in some previous works with an insufficient number of particles, the disc doubled its thickness during the period of simulation (e.g., \citealt{MartinezValpuesta_etal2006}, their figure~3) and it is not the physical effect but the consequence of an inadequate numerical modelling. To understand how relaxation affects the X-structures, we simulated one benchmark model with a different number of particles in the disc. We  measured time evolution of the disc thickness and the opening angle for each representation and compared it with results of previous studies.  We obtain that, starting from a sufficiently large number of particles in the disc ($N>10^6$), the X-structure appearance and morphology become independent of the number of particles. For models with different physical parameters, we use an even greater number of particles in the disc to achieve a better opening angle resolution ($1^\circ$--$2^\circ$) which allows us to compare modelled X-structures between themselves.  
\par
We have obtained that opening angles lie in the range from 25$^\circ$ to 42$^\circ$ throughout the entire evolution for all models. These values are consistent with observational data if we take into account the projection effect when a bar is viewed from different position angles. Resulting values of opening angles demonstrate a clear dependence on initial parameters of a galaxy model. The greater the relative mass of a dark halo, the smaller the opening angle. For an initial disc thickness and a Toomre parameter, we found that an increase of values results in an increase of the opening angle and changes the morphology of X-structures. The presence of a bulge prevents buckling at late stages of the disc evolution and results in very symmetric X-structures. The greatest variation in opening angles is obtained by varying the mass of a dark halo, while all other parameters lead only to smaller variations.
\par
The rest of the paper is organised as follows.
In Section~2 we give an overall description of numerical models, including their initial parameters and general evolution with time. 
In Section~3 we introduce an algorithm for measuring parameters of the X-structure and apply it to determine the effects of numerical relaxation. 
In Section~4, we use our algorithm to study the time evolution of the main parameters of X-structures in models with different halo contributions, the bulge concentration, the initial disc thickness and the Toomre parameter $Q$.
Section 5 discusses how the values of opening angles of simulated galaxies are consistent with the results of observations and previous numerical studies of the vertical bar evolution.
Finally, conclusions are presented in Section~6.
\par
\begin{table*}
\centering
\caption{Parameters of models}
\begin{tabular}{| c | c | c | c | c | c | c | c |c |}
\hline
Varied parameter &$M_\mathrm{h} (r < 4R_\mathrm{d})$& $M_\mathrm{b}$&$r_\mathrm{b}$& $ N_\mathrm{d} \cdot 10^6$ &$N_\mathrm{h} \cdot 10^6$&  $V_\mathrm{max}$, km $\cdot$ s$^{-1}$ & $z_d / R_\mathrm{d}$ & $Q$\\
\hline 
\hline
Halo mass & $1.0$ & --- & --- & 4 & 4& 194 & 0.05 & 1.2\\
\hline
---&$1.5$ & --- & --- & 4$^*$ & 4.5$^*$ & 211& 0.05 & 1.2\\   
\hline
--- & $2.25$ & --- & ---& 4 & 6.75 &  235& 0.05 & 1.2\\
\hline
--- &$3.0$ & --- & --- & 4 & 9 & 258& 0.05& 1.2 \\
\hline
\hline
Disc thickness &$1.5$ & --- & --- & 4 & 4.5 & 211 & 0.1& 1.2 \\
\hline
--- &$1.5$ & --- & --- & 4 & 4.5 & 211 & 0.2& 1.2 \\
\hline
\hline
Toomre parameter in a thin disc &$1.5$ & --- & --- & 4 & 4.5 & 211 & 0.05& 1.6 \\
\hline
--- &$1.5$ & --- & --- & 4 & 4.5 & 211 & 0.05 & 2.0 \\
\hline
\hline
Toomre parameter in a thick disc &$1.5$ & --- & --- & 4 & 4.5 & 211 & 0.1& 1.6 \\
\hline
--- &$1.5$ & --- & --- & 4 & 4.5 & 211 & 0.1& 2.0 \\
\hline
\hline
Central concentration &$1.5$ & 0.2 & 0.2 & 4 & 4.5 & 221 & 0.05& 1.2 \\
\hline
--- &$1.5$ & 0.4 & 0.2 & 4 & 4.5 & 266 & 0.05& 1.2 \\
\hline
--- &$1.5$ & 0.2 & 0.4 & 4 & 4.5 & 220 & 0.05& 1.2 \\
\hline
\multicolumn{9}{p{0.8\textwidth}}
{\footnotesize{\textit{Notes}: the first column gives a name of varied parameter in the set of models, each column starting from the second represents parameters of the models, one model on one line: $M_\mathrm{h}(R < 4R_\mathrm{d})$ is the mass of the halo within a sphere with radius $R=4R_\mathrm{d}$, where $R_\mathrm{d}$ is the scale length of the disc, $M_\mathrm{b}$ is total mass of the bulge, zero value means models without bulge, $N_\mathrm{d}$ and $N_\mathrm{h}$ are numbers of particles, which represent the disc and the halo, $V_\mathrm{max}$ is the value of maximal circular velocity if the disc scale length is equal to $3.5$ kpc, $z_\mathrm{d}/R_\mathrm{d}$ is the initial ratio of the disc vertical scale length to the disc scale length. Double horizontal lines separate different sets of models. Dashes in $M_\mathrm{b}$ and $r_\mathrm{b}$ represent models without bulge. Symbol "$^*$" denotes a fiducial model simulated with different number of particles (see Table~\ref{tab:models_n})}.}
\end{tabular}
\label{tab:models_mu}
\end{table*} 

\begin{figure*}
\begin{minipage}[t]{0.3 \textwidth}
\includegraphics[scale=0.3]{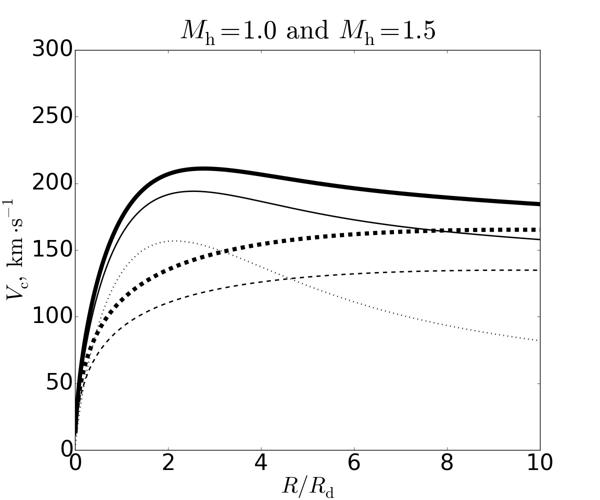}
\end{minipage}%
\hspace{0.5cm}
\begin{minipage}[t]{0.3 \textwidth}
\includegraphics[scale=0.3]{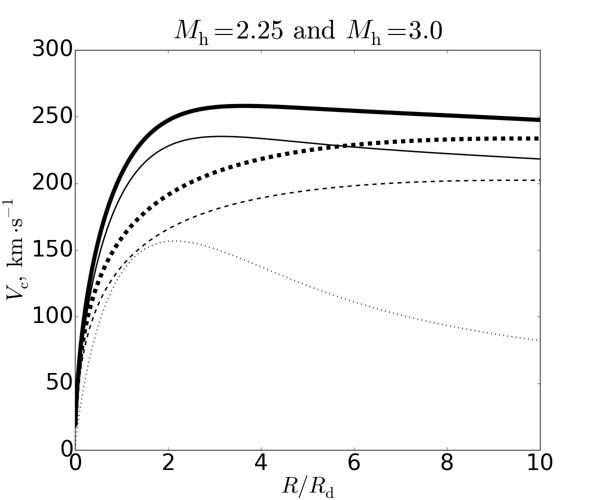}
\end{minipage}
\hspace{0.5cm}
\begin{minipage}[t]{0.3 \textwidth}
\includegraphics[scale=0.3]{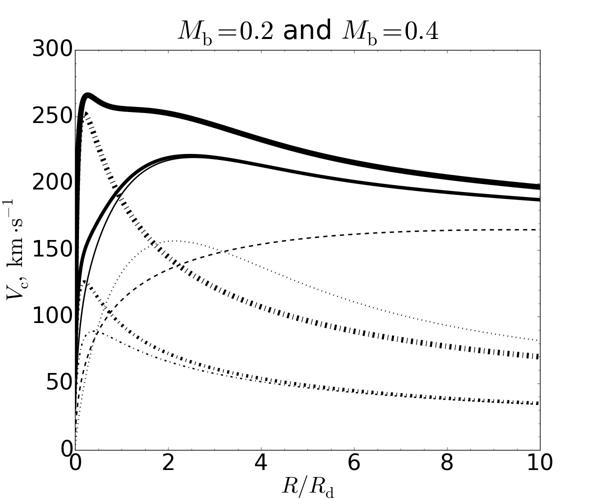}
\end{minipage}
\caption{The rotation curves of the galaxy models with different halo (left and middle panels) and bulge contributions (right panel). The dotted line corresponds to the disc contribution. Dashed lines correspond to the halo contribution. \textit{Left}: $M_\mathrm{h}=1.0$ (thin lines) and $M_\mathrm{h}=1.5$ (thick lines); \textit{Middle}: $M_\mathrm{h}=2.25$ (thin lines) and $M_\mathrm{h}=3.0$ (thick lines); \textit{Right}: $M_\mathrm{h}=1.5$ and different bulges: a light and diffuse bulge with $M_\mathrm{b}=0.2$ and $r_\mathrm{b}=0.4$ (a thin solid line), an intermediate bulge with $M_\mathrm{b}=0.2$ and $r_\mathrm{b}=0.2$ (a solid line of intermediate thickness), a heavy bulge with $M_\mathrm{b}=0.4$ and $r_\mathrm{b}=0.2$ (a thick solid line). Dashed-dotted lines of different thicknesses correspond to the different bulge contributions in a similar manner as solid lines.}
\label{fig:vel_curves}
\end{figure*}

\begin{table}
\centering
\caption{Parameters of models with the same mass of the dark halo $M_\mathrm{h}=1.5$ and different number of particles in the disc and the halo}
\begin{tabular}{| c | c | c | c | c |}
\hline
$ N_\mathrm{d} \cdot 10^6$ &   $N_\mathrm{h} \cdot 10^6$& $m_\mathrm{h}/m_\mathrm{d}$ & $\varepsilon_\mathrm{d}/R_\mathrm{d} \cdot 10^{-2}$ & $\varepsilon_\mathrm{h}/R_\mathrm{d}  \cdot 10^{-2}$ \\
\hline 
\hline
 0.2 & 1.2 & 1.14 & 1.0 & 2.0 \\ 
\hline 
1 & 4.5 & 1.8 & 0.5 & 1.2 \\
\hline 
1 & 9 &0.9& 0.5 & 1.0 \\
\hline 
2 & 9 &1.8& 0.47 & 1.02  \\
\hline 
4 & 4.5 &7.3& 0.37 & 1.29  \\
\hline
8 & 9 &7.3& 0.29 & 1.02 \\
\hline          
\multicolumn{5}{p{0.45\textwidth}}
{\footnotesize{\textit{Notes}: each column represents parameters of the models, one model on one line: $N_\mathrm{d}$ and $N_\mathrm{h}$ are numbers of particles which represent the disc and the halo, $m_\mathrm{h}/m_\mathrm{d}$ is the ratio of the mass of the particles constituted the halo to the mass of the particles constituted the disc, $\varepsilon_\mathrm{d}$ and $\varepsilon_\mathrm{h}$ are softening lengths of the disc and the halo.}}
\label{tab:models_n}
\end{tabular}
\end{table}

\section{Modelling of X-structures}

\subsection{Numerical model}
To investigate the dependence of X-structure properties on characteristics of modelled galaxies, we consider thirteen different three-dimensional $N$-body models of galaxies with different physical parameters (see Table~\ref{tab:models_mu}) and six models with the same physical parameters but represented by a different number of particles (Table~ \ref{tab:models_n}). Each model comprises a stellar disc and a dark spherical halo. Three models, which were intended for studying the effect of central concentration, also include a bulge component. All models share similar density profiles for each component specified below. 
\par
The disc density profile is exponential in the radial direction and consists of isothermal sheets in the vertical direction:
\begin{equation}
\rho(r,z) = \frac{M_\mathrm{d}}{4\pi R_\mathrm{d}^2 z_\mathrm{d}} \cdot \exp(-R/R_\mathrm{d}) \cdot \sech^2(z/z_\mathrm{d}) \,,
\label{eq:sigma_disc} 
\end{equation}
where $R_\mathrm{d}$ is the disc radial scale length, $z_\mathrm{d}$ is the disc scale height, $M_\mathrm{d}$ is disc mass. 
\par
The halo has a truncated two-power density profile:
\begin{equation}
\rho = 
\frac{C_\mathrm{h}\,T(r/r_\mathrm{t})}
{(r/r_\mathrm{s})^{\gamma_0}
\left((r/r_\mathrm{s})^{\eta}+1\right)^
{(\gamma_{\infty}-\gamma_0)/\eta}} \,,
\label{eq:NFW}
\end{equation}
where $r_\mathrm{s}$ is the halo scale radius, $r_\mathrm{t}$ is the halo truncation radius, $\eta$ is the halo transition exponent, $\gamma_0$ is the halo inner logarithmic density slope, $\gamma_{\infty}$ is the halo outer logarithmic density slope, $C_\mathrm{h}$ is the parameter defining the full mass of the halo $M_\mathrm{h}$ and $T(x)$ is the truncation function: 
\begin{equation}
T(x) = \frac{2}{\sech{x} + 1/\sech{x}} \,.
\end{equation}
We use $\eta=4/9$, $\gamma_0=7/9$, $\gamma_{\infty}=31/9$, which makes the halo density profile to be very similar to the Navarro-Frenk-White profile (\citealt*{NFW}) with a slightly steeper slope in the inner region.
\par
The bulge is defined by \cite{Hernquist1990} density profile:
\begin{equation}
\rho_\mathrm{b} = \frac{M_\mathrm{b}\, r_\mathrm{b}}{2\pi\,r\,(r_\mathrm{b} + r)^3} \,, 
\end{equation} 
where $r_\mathrm{b}$ is the scale parameter and $M_\mathrm{b}$ is the total bulge mass. 
In simulations, each component is represented by a set of particles denoted by $N_\mathrm{d}$ for the disc, $N_\mathrm{h}$ for the halo and $N_\mathrm{b}$ for the bulge. The disc, the halo and the bulge  are self-consistent, i.e. all components evolve under the influence of their mutual gravitational field.
\par 
Disc parameters are fixed by the choice of the system of units: $M_\mathrm{d}=1, \, R_\mathrm{d}=1$. For simplicity, we also set $G=1$. If one uses $R_\mathrm{d}=3.5$ kpc and $M_\mathrm{d}=10^{10} M_{\sun}$ then the time unit will be $t_\mathrm{u} = 13.22$ Myr. \textcolor{black}{
Hereafter, unless otherwise specified, when referring to a variable measured in units of length we will address to the variable itself without notation of units, which means that the variable is measured in simulation length units ($1$ length unit = $3.5$ kpc).
}
\par 
The halo scale length and the truncation radius are the same for all models: $r_\mathrm{s}=6$, $r_\mathrm{t}=15$.
\par 
The initial velocity dispersion profile $\sigma_R$ for the disc is exponential:
\begin{equation}
\sigma_R = \sigma_0 \cdot \exp(-R/2 R_\mathrm{d}),
\end{equation}
where $\sigma_0$ is defined by the choice of the value of the Toomre parameter $Q_{\mathrm{T}}(R_\sigma)=Q_0$ at $R_\sigma=2\,R_\mathrm{d}$. Further, we characterise the degree of disc heating by the value of $Q_0$.
\par
For convenience we grouped all models according to their physical parameters.
We obtained five different sets. Values of parameters which vary from one set to another are presented in Table~\ref{tab:models_mu}. 
\par
The first set is devoted to explore the influence of the dark halo mass on the shape of the X-structure. Varying the constant $C_\mathrm{h}$ in the density profile of the dark halo Eq.~\eqref{eq:NFW}, we obtain models with different contributions of the dark halo to the overall gravitational field. Different models are characterised by the relative contribution of the dark halo to the overall mass within the stellar disc $\mu = M_\mathrm{h}(R<4 R_\mathrm{d}) / M_\mathrm{d}$, where $M_\mathrm{h}(R<4 R_d)$ is the mass of the halo within $4 R_\mathrm{d}$. In our system of units $\mu$ equal to $M_\mathrm{h}(R<4R_\mathrm{d})$ and hereafter we refer to this quantity as $M_\mathrm{h}$. We constructed four models with the following values of $M_\mathrm{h}: \,1.0\,, 1.5, \, 2.25, \, 3.0$ (see Table~\ref{tab:models_mu}). Such a choice of dark halo mass values is motivated by observational data. The values considered are quite typical dark halo masses within the optical radius of a stellar disc \citep{deBlok_McGaugh1997,Khoperskov2002,Pizagno_etal2005,Bizyaev_Mitronova2009}. They are inherent in high surface brightness galaxies falling within the range  $1 \lesssim \mu \lesssim 3$ (see \citealt{deBlok_McGaugh1997}). As to the X-structures, we find them, as a rule, in bright galaxies \citep{Laurikainen_Salo2017,Savchenko_etal2017}. The rotation curves of the models with the contributions of individual subsystems are shown in Fig.~\ref{fig:vel_curves}.
\par
All other sets of models are obtained by varying parameters which entering into criteria of bending or in-plane instabilities: the disc thickness ($z_{\mathrm{d}}$), the mass concentration ($M_{\mathrm{b}}$ and $r_{\mathrm{b}}$) and the degree of stellar disc heating ($Q_{\mathrm{T}}$). 
Disc thickness is varied between  $z_{\mathrm{d}}=0.05$  (very thin disc) and $z_{\mathrm{d}}=0.2$ (thick disc). Mass of the bulge is varied from $M_\mathrm{b}= 0.0$ (bulgeless model) to $M_\mathrm{b}= 0.4$ (heavy bulge) and the bulge scale length is varied from $r_\mathrm{b}=0.2$ to $r_\mathrm{b}=0.4$. The Toomre parameter $Q_\mathrm{0}$ is varied from $1.2$ (cool disc) to $2.0$ (hot disc). 
\par
Prior to any simulation, which is used to determine physical dependencies, we investigate how possible numerical effects influence X-structure properties. For this purpose, we run simulations of a fiducial model with a moderate halo, $M_\mathrm{h}=1.5$, and a thin disc, $z_d=0.05$, without a bulge, which is represented by different numbers of particles in the disc and the halo (this model is similar to the bulgeless Milky Way with a slightly heavier dark halo (\citealt{Binney_Tremaine2008},hereafter BT, pp 113-117, especially Tab 2.3 on p. 113; \citealt{Piffl_etal2014}). The considered range of the number of particles covers almost two orders of magnitude from $N_\mathrm{d} = 2 \cdot 10^5, \, N_\mathrm{h} = 1.2 \cdot 10^6$ to $N_\mathrm{d} = 8 \cdot 10^6, \, N_\mathrm{h} = 9 \cdot 10^6$ (Table~\ref{tab:models_n}).
\par 
As argued in Sec.~\ref{sec:measure_x}, for our procedure of measuring X-structure parameters, one must use at least $10^6$ particles in the disc and $4.5 \cdot 10^6$ in the halo. Taking into account the results of the relaxation analysis and the analysis of the evolution of opening angles described in Sec.~\ref{sec:model_des} and \ref{sec:measure_x}, we choose an even greater number of particles to represent each model. As a rule, models with different physical parameters are populated with $4\cdot 10^6$ disc particles and $4.5 \cdot 10^6$ halo particles. For dark halos with $M_\mathrm{h} >1.5$, the number of particles $N_\mathrm{h}$ is varied in such a way that the mass of each halo particle remains unchanged from one model to another. For a model with a light halo $M_\mathrm{h} \,= \, 1.0$, we use $4 \cdot 10^6$ halo particles. If bulge particles present they have the same mass as disc particles, and this condition determines the overall number of particles in the bulge.
\par 
The initial equilibrium state is prepared via a script for constructing the equilibrium multicomponent model of a galaxy {\tt{mkgalaxy}} \citep{McMillan_Dehnen2007} from the toolbox for N-body simulation {\tt{NEMO}} \citep{Teuben_1995}. 
\textcolor{black}{
Both spherical components were assumed to have an isotropic velocity distribution. This assumption has the advantage that the distribution function (DF) is exact \citep{Eddington1916}. For a disc the DF for dynamically warm thin stellar discs, which depends on three integral of motions and approximately reproduces predescribed surface density and velocity dispersion profiles, is used \citep{Dehnen1999}.
} 
On the first step, spherical initial conditions for the halo and the bulge in the presence of the monopole part of the disc potential as well as the external potential are generated. Then both spheroids are adjusted to the presence of the full disc potential, rather than only its monopole counterpart. The final step is to populate the disc component. In this step, the potential of the halo and the bulge in their turn are considered as external. The smoothed, azimuthally averaged potentials of these components are described by potential expansion.
\par
The gravitational softening length $\varepsilon$ is different for each subsystem and scaled in accordance with the number of particles in the disc and the halo:
\begin{equation}
\varepsilon_\mathrm{d;\,h} = \varepsilon_\mathrm{0,\, d;\,h} \cdot \left(\frac{N_\mathrm{0,\, d;\,h}}{N_\mathrm{d;\,h}}\right)^{1/3} \,,
\end{equation} 
where $\varepsilon_\mathrm{0,\,d}=0.01$ for the number particles in the disc $N_\mathrm{d,\,0}=2 \cdot 10^5$ and $\varepsilon_\mathrm{0,\,h}=0.02$ for the number particles in the halo $N_\mathrm{0,\, h}=1.2 \cdot 10^6$. For the bulge, the softening length coincides with that of the disc.
\par
We follow the evolution of modelled galaxies using the fastest $N$-body code for one CPU {\tt{gyrfalcON}} \citep{Dehnen2002}. This code is a realisation of a tree-code algorithm, for which close iterations are calculated straightforwardly from Newton law (particle-particle interaction) and distant interactions are approximated by cell-particle interaction. A minimal size of cells is defined by the tolerance parameter $\theta$.  We use $\theta=0.6$, which gives a satisfactory ratio between the time of integration and the accuracy of integration for models with flat components \citep{Dehnen2002}. We vary the integration time step according to the rule $0.1\sqrt{\varepsilon/|\vec a|}$, where $\vec a$ is the gravitational acceleration.

\subsection{General description of model evolution} \label{sec:model_des}
\begin{figure*}
\includegraphics[scale=0.31]{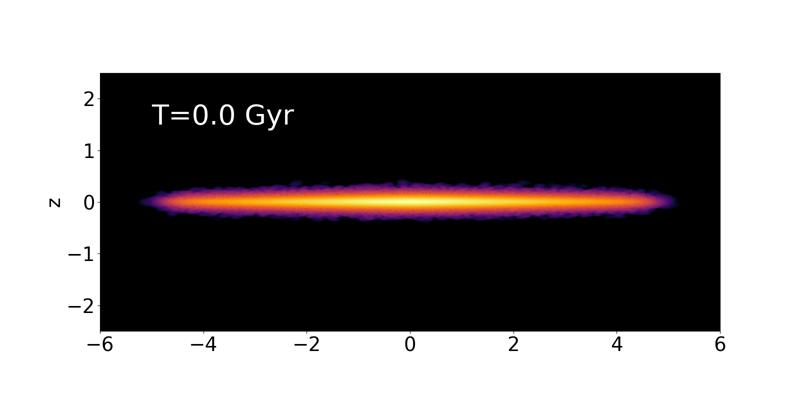}
\includegraphics[scale=0.31]{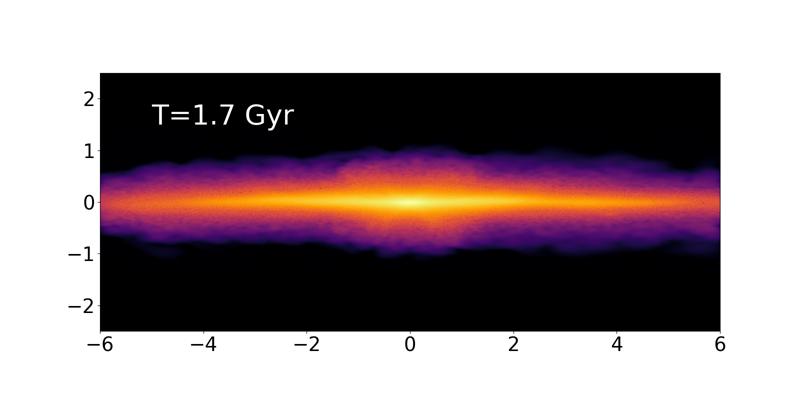}
\includegraphics[scale=0.31]{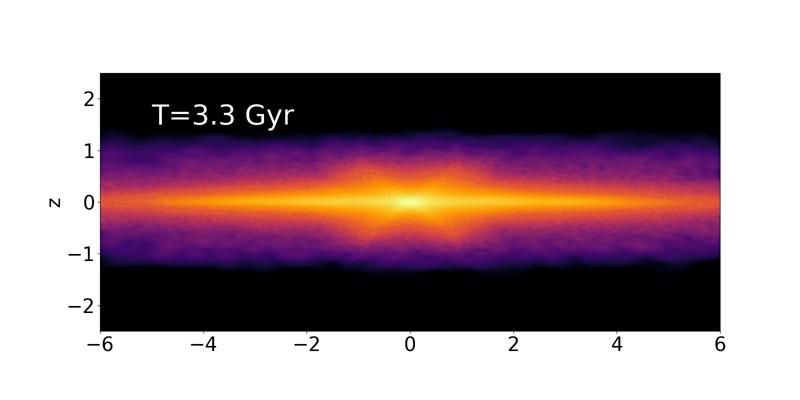}
\includegraphics[scale=0.31]{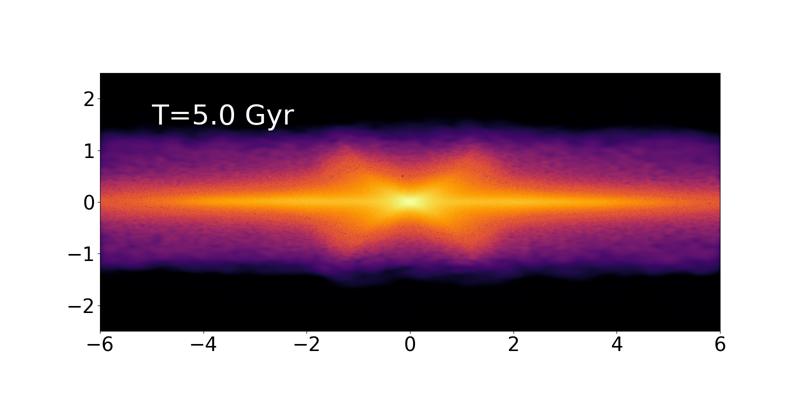}
\includegraphics[scale=0.31]{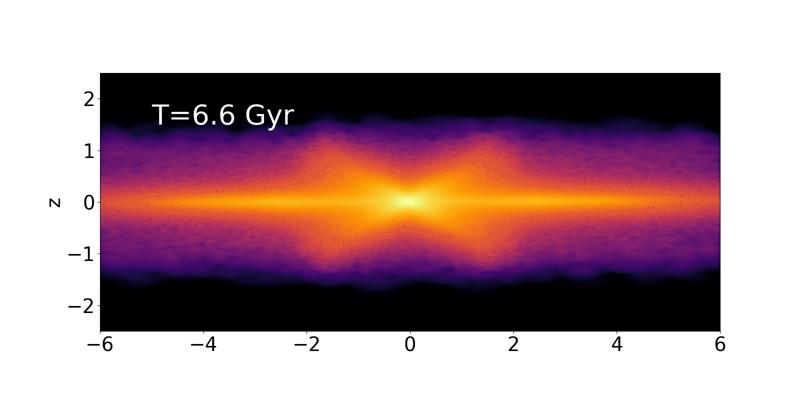}
\includegraphics[scale=0.31]{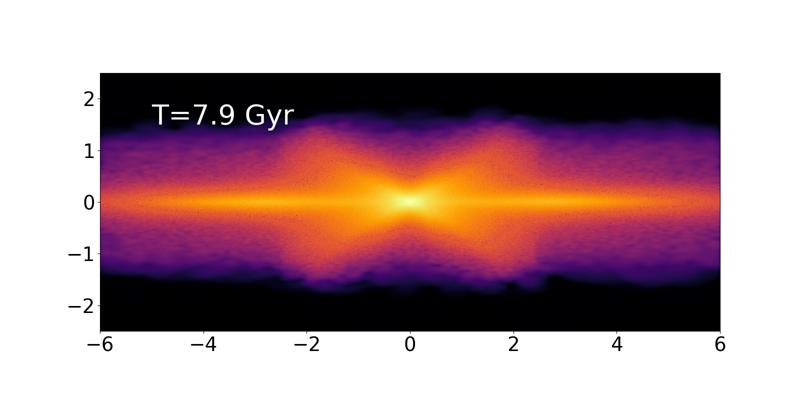}
\caption{The time evolution of the model with $M_\mathrm{h}=1.5$ and the highest spatial resolution, $N_\mathrm{d} = 8 \cdot 10^6$ , $N_\mathrm{h}=9 \cdot 10^6$, ($x-z$) projection, 
\textcolor{black}{displayed in simulation units}.}
\label{fig:snapshots_xz}
\end{figure*}

\begin{figure*}
\begin{minipage}[t]{0.32\textwidth}%
\includegraphics[scale=0.2]{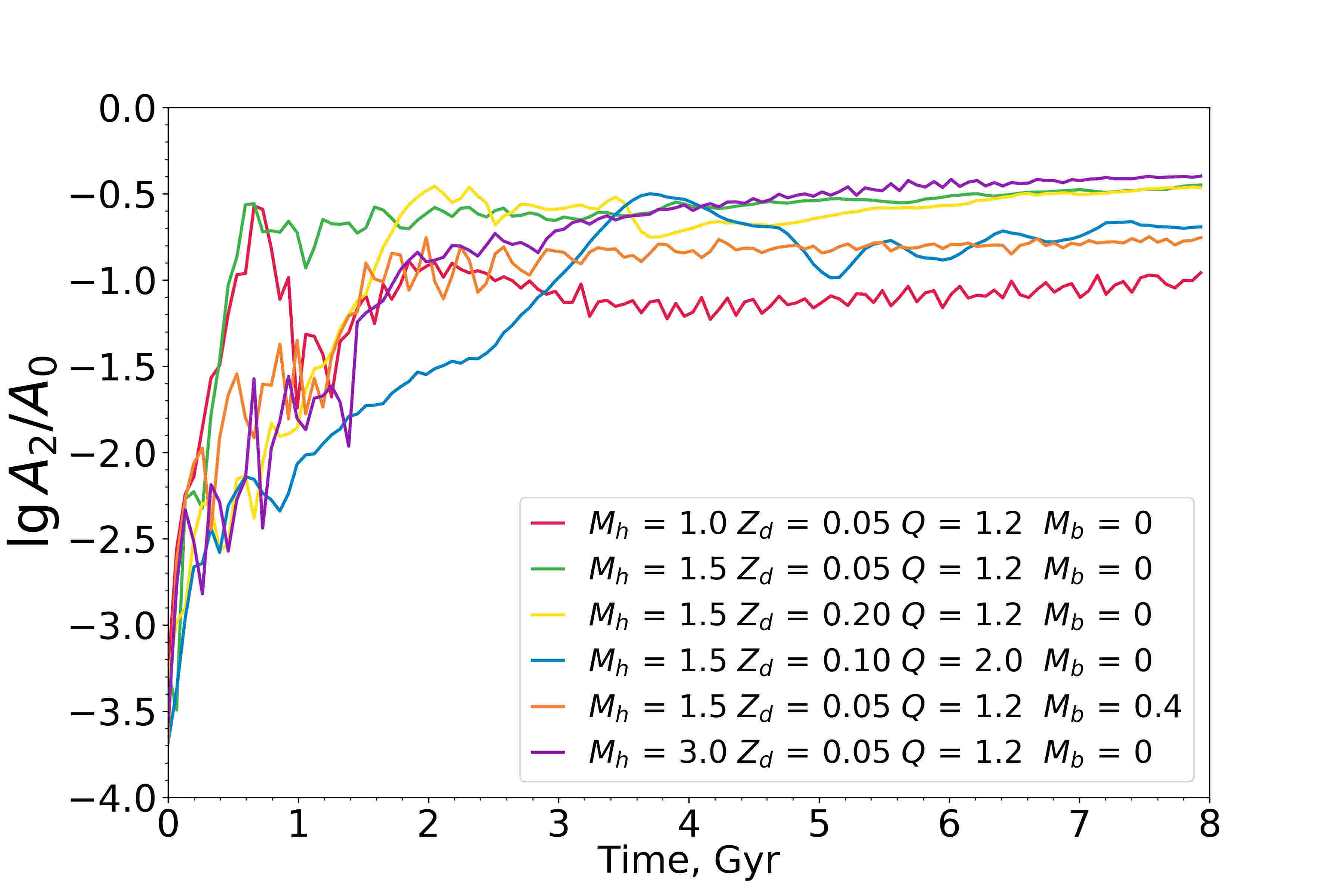}
\end{minipage}%
\begin{minipage}[t]{0.32\textwidth}%
\includegraphics[scale=0.096]{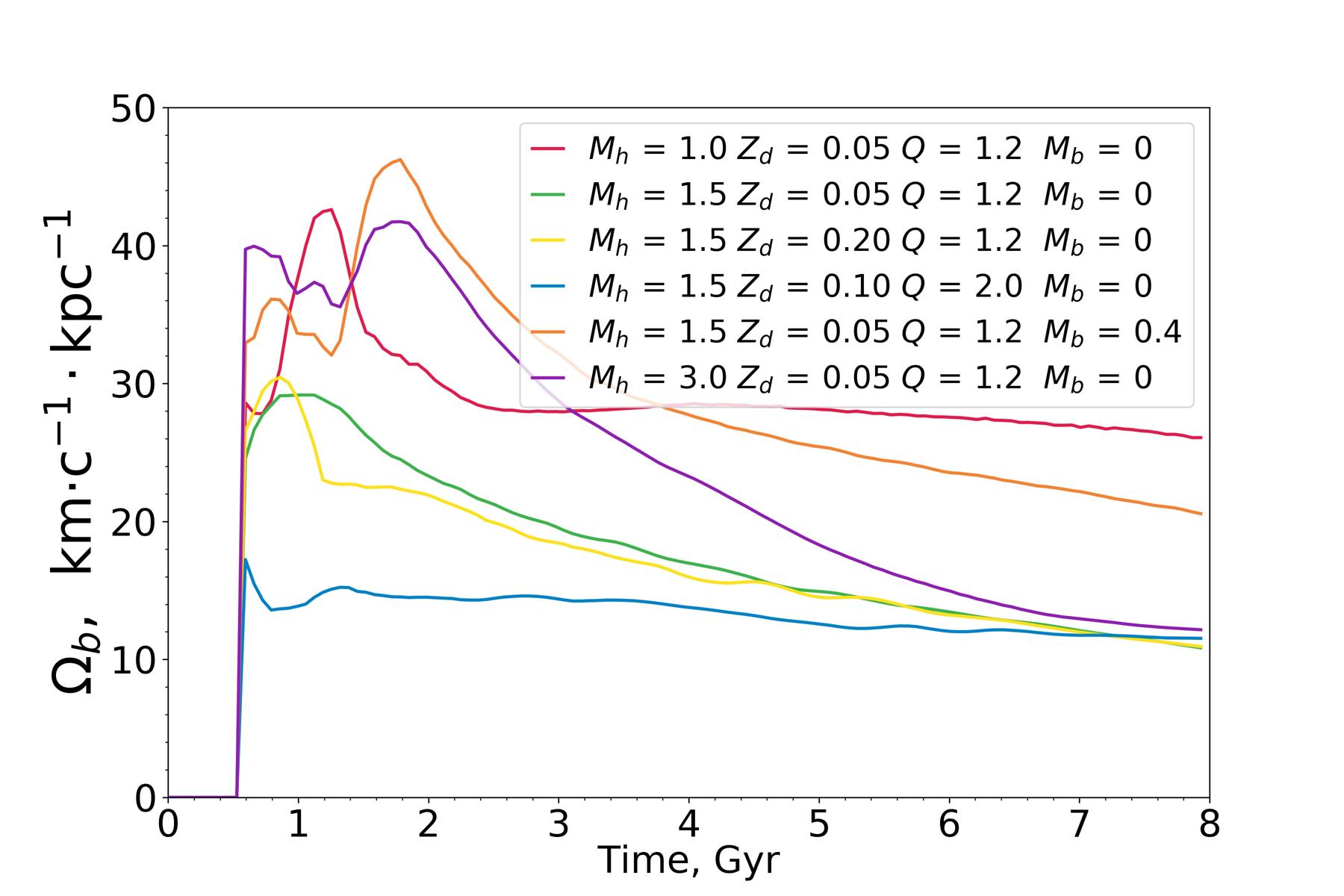}
\end{minipage}%
\begin{minipage}[t]{0.32\textwidth}%
\includegraphics[scale=0.2]{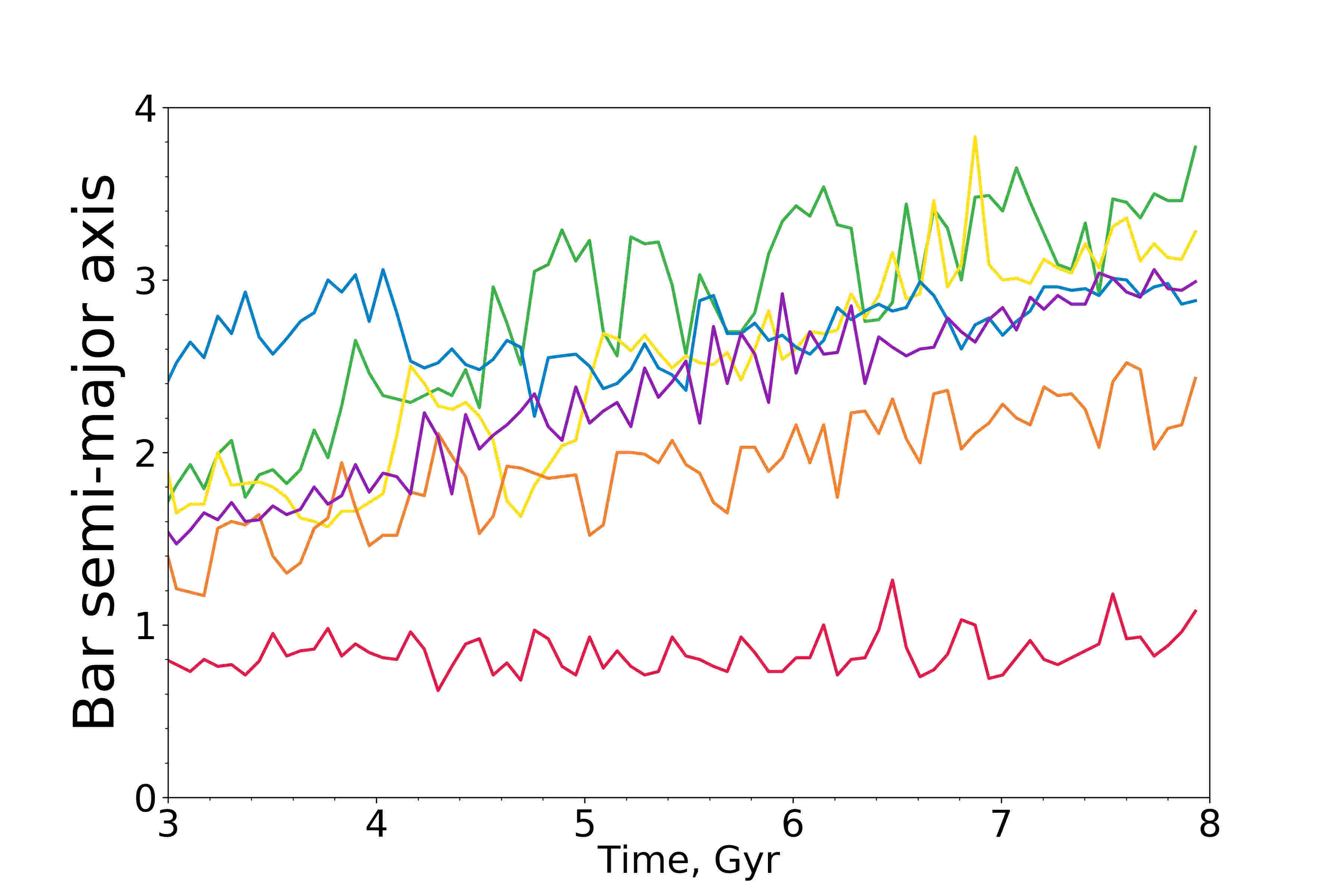}
\end{minipage}%
\caption{The logarithm of the amplitude (left), the pattern speed (middle) and the length of the semi-major axis (right) of a bar for models with boundary values of parameters from Table~\ref{tab:models_mu} and a fiducial bulgeless model with $M_\mathrm{h}=1.5$~, $Z_\mathrm{d}=0.05$,~$Q=1.2$.}
\label{fig:ampl}
\end{figure*}

All models with different physical parameters from Table~\ref{tab:models_mu} are unstable with respect to the bar instability and over time, the bar grows in the vertical direction (Fig.~\ref{fig:snapshots_xz}). Consequently, despite a significant difference in initial conditions, the evolutionary path of all models can be naturally described in the following way based on the state of a bar. There are three different states:\\
1) \textit{Before a bar forms}. At this stage, the whole system is also the subject to the vigorous spiral instability. As a rule, there are more than 4-5 spiral waves of different wavelengths (short and long waves present in conjunction). At some point in time spiral waves tend to slow down its fast-changing pattern and form a small and fast rotating bar in the centre of the system, which marks the end of this stage. A characteristic time of this stage coincides with bar formation time, which can be roughly estimated \textcolor{black}{
as the stabilisation time of the second azimuthal Fourier harmonic amplitude $A_2$: 
\begin{equation}
A_2 = \int \rho(R, \phi, z) \exp \left\{ 2i\phi \right\} R \, \, dR \, d\phi \, dz.
\end{equation}
}
In our models, the characteristic period of bar formation is about $1$ Gyr for cool discs and $2$ Gyrs for hot discs (see Fig.~\ref{fig:ampl}).\\
2) \textit{A formed bar grows and slows down}.
To trace further evolution of a bar, we calculated a pattern speed and a half-length of its major axis (Fig.~\ref{fig:ampl}). 
\par 
The pattern speed is calculated as the derivative of the position angle divided by the azimuthal wave number \mbox{$m=2$}: \mbox{$\Omega_b= d\psi/(2\cdot dt)$}, while the position angle is calculated via Fourier-transform of the density of the disc:
\mbox{$\psi=\arctan\left(\Im(A_2) / \Re(A_2)\right)$.}
\par
\begin{figure}
\centering
\includegraphics[scale=0.40]{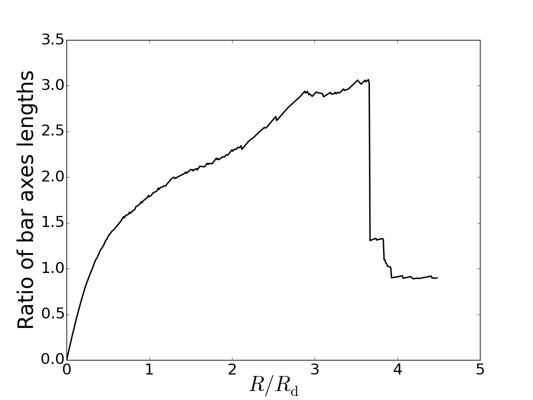}
\caption{The ratio of lengths of axes of elliptical isophotes in the fiducial model at the time moment $t=570$ units (7.5~Gyrs). The point where the ellipticity suddenly drops down is assumed to be the half-length of the major axis of the bar.}
\label{fig:ellipticity}
\end{figure}

 
The length of the bar major axis can be defined as the distance along the bar major axis from the centre to the point where the ellipticity of isophotes abruptly drops to zero  (see Fig. ~\ref{fig:ellipticity}). To simplify the analysis we do not fit elliptical isophotes but calculate the density distribution along major and minor axes and assume that the ratio of distances along these axes to equal densities is an equivalent to the axis ratio of isophotes covering the bar. 
\par
As shown in Fig.~\ref{fig:ampl}, a formed bar rapidly decreases the pattern speed and, at the same time, grows in size, which is consistent with previous studies of bars \citep{WeibergThesis, Debattista1998,Athanassoula2003,Valenzuela_Klypin2003}. At the end of simulations, the typical pattern speed is about $15-30$ km$\cdot$c$^{-1}$kpc$^{-1}$ and semi-major axis of a bar is about 2-3 times of the initial disc scale length. 
\par
While a bar grows in a disc plane, it also starts to thicken in the vertical direction. For models with light halos ($M_\mathrm{h}=1.0$ and $M_\mathrm{h}=1.5$) the bar jumps out of the disc plane almost instantly as its forms while for models with heavy halos ($M_\mathrm{h}=2.25$ and $M_\mathrm{h}=3.0$) there is a period of time of about $1$ Gyr when a bar preserves its plane morphology and only after that starts to thicken. 
\par
3) \textit{Thickening of a bar}.
As it was proved in many numerical (e.g. \citealp{Raha_etal1991,Merritt_Sellwood1994,Sotnikova_Rodionov2003}) and theoretical studies, a bar appears to be a subject to the bending instability of fire-horse type \citep{Toomre1966,Poliachenko_Shukhman1977,Araki1985}, which is often called a buckling  instability. By all means, this instability leads to bar thickening and, consequently, to formation of X-shaped structures. In our models, all bars ended up with clear X-structures, like those shown in Fig.~\ref{fig:snapshots_xz}. Newly born X-structures are short and have a large opening angle. As they evolve, they grow in length and their opening angle decreases. 
\par
Almost all models undergo an additional buckling, which manifests itself in a repeated vertical asymmetry of the X-structure. Visually, for such X-structures, some rays are significantly longer than the others (Figs.~\ref{fig:im_thickness}, \ref{fig:im_toomre} and \ref{fig:im_toomre2}). Moreover, in some models due to so strong but temporal buckling, \textit{double} X-structures can form (see Fig.~\ref{fig:im_toomre}) and  secondary four rays, coming in time with a symmetrical state, do not emerge from the centre of the disc. Obviously, such a buckling can have a great impact on the morphology of X-structures. Therefore, we postpone a detailed description of the buckling history of each model to Sec.~\ref{sec:sec_buckling}, where we analyse it in conjunction with the evolution of X-structure parameters. 

\subsection{Evolution of disc thickness}

\begin{figure*}
\begin{minipage}[t]{0.49\textwidth}%
\includegraphics[scale=0.15]{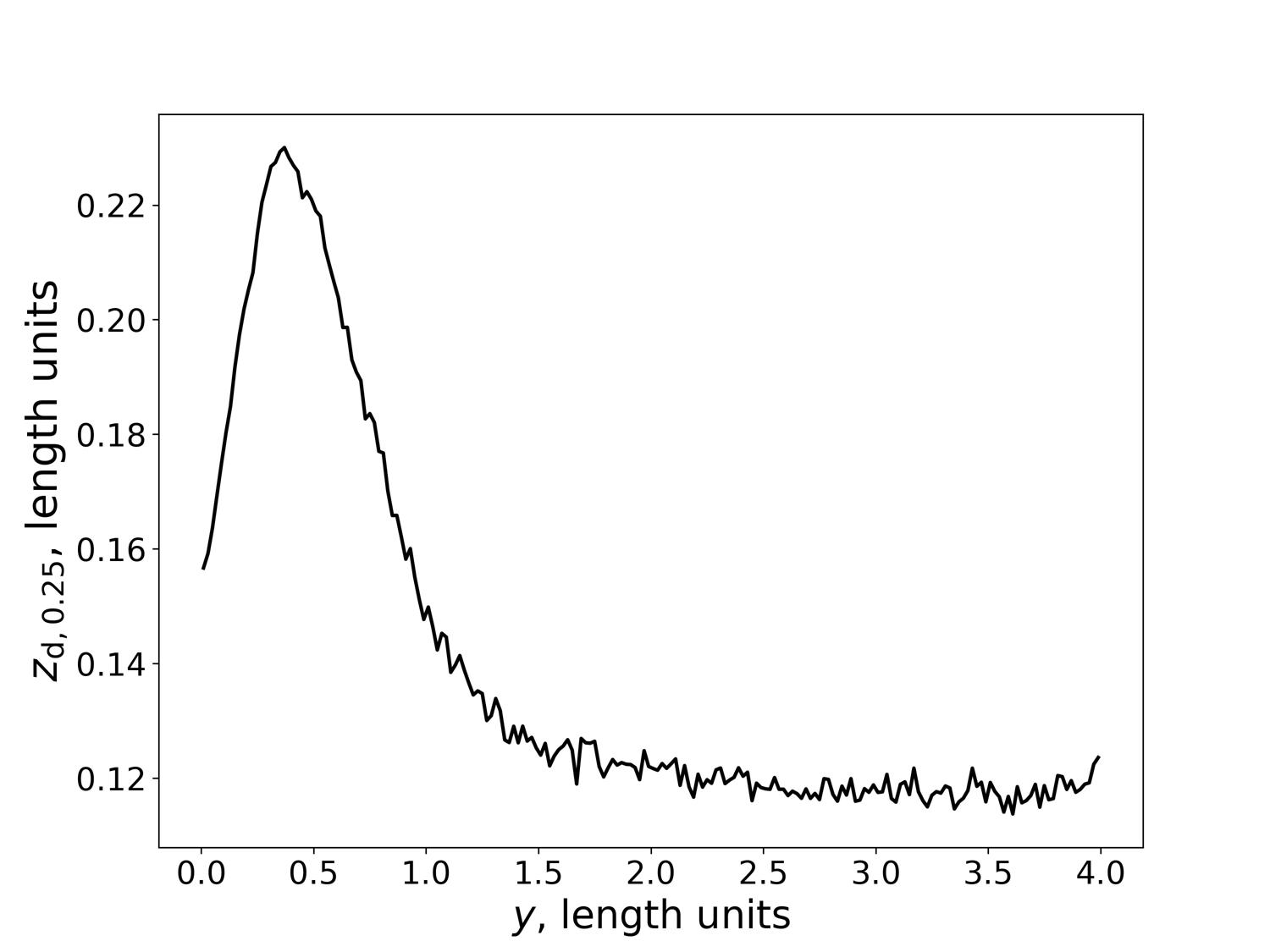}
\end{minipage}
\hfill
\begin{minipage}[t]{0.49\textwidth}%
\includegraphics[scale=0.24]{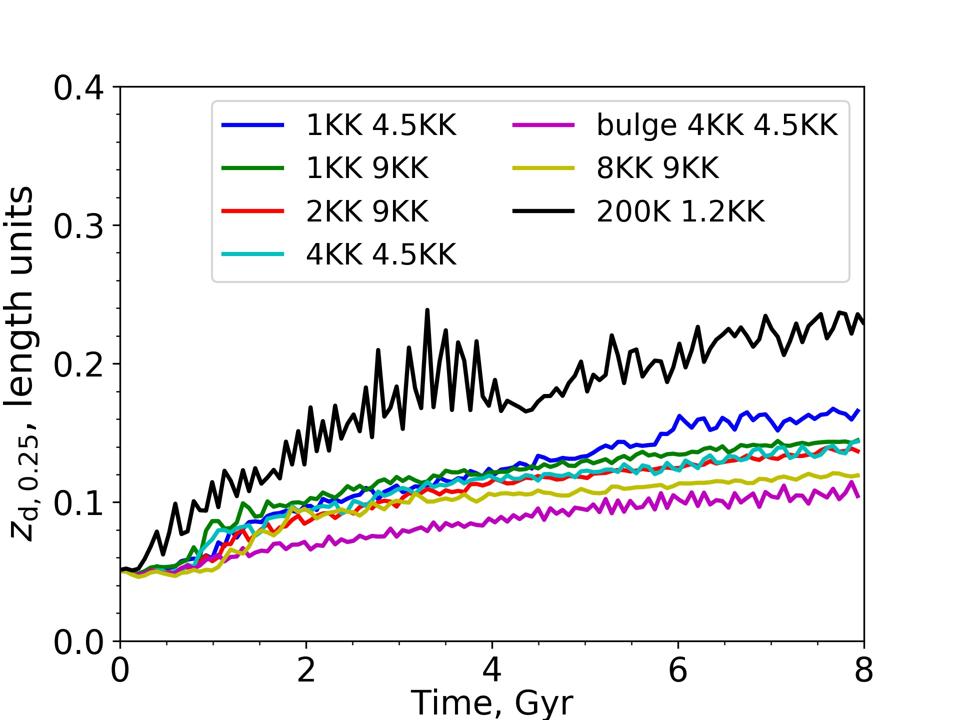}
\end{minipage}%
\caption{The disc thickness. \textit{Left}: The typical dependence of the disc vertical scale length $z_\mathrm{d, 0.25}$ at the distance to the disc centre $y$ calculated in $(y-z)$ projection for the model with $M_\mathrm{h}=1.5$ at the time moment $t=7.5$ Gyr (570 time units). \textit{Right}: The evolution of the disc vertical scale length $z_\mathrm{d, 0.25}$ (see text for details) for the models with $M_\mathrm{h}=1.5$ and different number of particles in the disc and the halo.}
\label{fig:relax}
\end{figure*}

\par
In numerical simulations of thin discs there is a problem associated with numerical relaxation in modelled galaxies. 
Observed galaxies contain $N\approx 10^{9} - 10^{11}$ stars. Typically, the modelled stellar disc is represented by $N=10^5-10^7$ particles. The specific period of time after which the velocity dispersion due to scattering becomes as large as the initial velocity is called a relaxation time. The relaxation time for real galaxies is greater than the Hubble time. For modelled galaxies it depends on the chosen number of particles (see BT, pp 34--37). For the scattering in the disc plane, the relaxation time can be suppressed by using number of particles $N > 10^5$. However, the situation is not nearly the same for the vertical direction.
Numerical studies of the vertical structure of discs     \citep{Sellwood2013,Rodionov_Sotnikova2013} proved that the relaxation in the vertical direction can not be suppressed even by using $N=10^6$ particles. It results in thickening of the stellar disc and in an increase of the vertical-to radial velocity dispersion ratio \citep{Sellwood2013,Rodionov_Sotnikova2013}. In some previous numerical studies, these effects can be clearly seen \citep{MartinezValpuesta_etal2006}.
\par 
The chosen number of particles in a disc and a halo affects the evolution of models in the vertical direction. In this work, we investigate how the choice of the number of particles influences the evolution of X-structure angles. We pursue two goals. First, to compare parameters of X-structures with observational data we have to be sure that a measured opening angle of an X-structure of modelled galaxies is independent of the number of particles which represent a stellar disc and a halo. Second, it is useful for further researches to determine how many particles are needed for accurate modelling of an X-structure.
\par
The influence of the relaxation (i.e. particle-particle scattering processes) in the vertical direction on stellar systems can be analysed in terms of the disc thickness \citep{Sellwood2013,Rodionov_Sotnikova2013}. We estimate the disc thickness in the following way. First, we rotate the model galaxy in such a way that the major axis of the bar will coincide with the $x$-axis at all times. Next, we calculate a projection density of the image of the modelled galaxy along the major axis of the bar ($y-z$ projection). We use a projection along the major axis of the bar to eliminate the X-structure contribution. The projection density is calculated on a square grid $(y_i, z_j)$:
\begin{equation}
y_i = y_\mathrm{min} + \Delta y \cdot i, \, z_j = z_\mathrm{min} + \Delta z \cdot j \,.
\label{eq:grid}
\end{equation}
Here we use the following values:
$\Delta y = \Delta z = 0.02$, 
$y_\mathrm{min} = -4$, $z_\mathrm{min} = -0.25$, $y_\mathrm{max}=4$,  $z_\mathrm{max}=0.25$,
$0\leq i \leq \displaystyle 
(y_\mathrm{max} - y_\mathrm{min})/\Delta y$, 
$0\leq j \leq \displaystyle (z_\mathrm{max} - z_\mathrm{min})/ \Delta z$. There are not enough particles to accurately estimate the disc density at high $z$ in models with the low spatial resolution. Therefore, we set the upper limit of the height $z_\mathrm{max} = 0.25$ to compare models with different number of particles. 
\par
We averaged the obtained values of the projection density over four quadrants. Next, we approximated the average dependence of the projection density on the height for each $y$-cut:
\begin{equation}
\rho_{yz} (y, z, t) = \rho_{0, yz}(y, t)\cdot \sech^2(z/z_\mathrm{d, 0.25}) \,.
\end{equation}
and used the method of the least squares to obtain the best fit value of $z_\mathrm{d, 0.25}$. Then we averaged found values over the interval  $3 \leq y \leq 4$, where the disc thickness is nearly constant (see Fig.~\ref{fig:relax}, left plot).
\par
As can be seen from Fig.~\ref{fig:relax} (right plot), the evolution of models is strongly dependent on the number of particles. For $N_d = 2 \cdot 10^5$, the thickness of the disc increases by a factor of five while for $N=8 \cdot 10^6$ it only doubles. Generally, more particles in the disc corresponds to a slower evolution of the disc thickness. The exception from this rule is the model with the bulge and $N_\mathrm{d} = 4 \cdot 10^6$. This model exhibits the lowest growth rate of the disc thickness, despite the smaller number of particles in the disc than in the model with the highest spatial resolution ($N_\mathrm{d}=8 \cdot 10^6$) and without the bulge. Such behaviour is consistent with the concept of the stabilising effect of the bulge 
on the disc thickness \citep{Sotnikova_Rodionov2005}. 
\par
As the chosen number of particles significantly changes the evolution of the vertical structure of considered models, we must study the dependence of the properties of X-structures on the number  particles to avoid numerical effects when we compare real observational data with models and modelled X-structures between themselves.

\section{The X-structure analysis} \label{sec:measure_x}

\subsection{Measuring parameters of X-structure}

\begin{figure*}
\begin{minipage}[t]{0.33\textwidth}%
\includegraphics[scale=0.3]{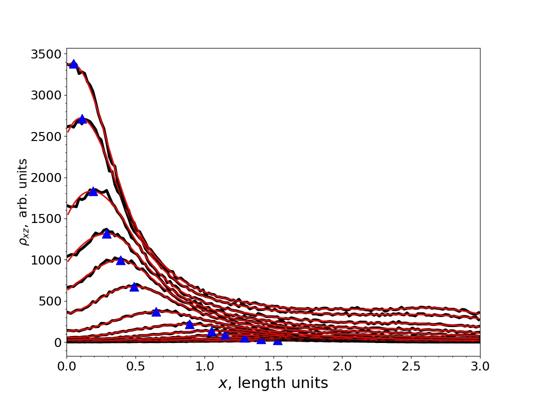}
\end{minipage}%
\begin{minipage}[t]{0.33\textwidth}%
\includegraphics[scale=0.3]{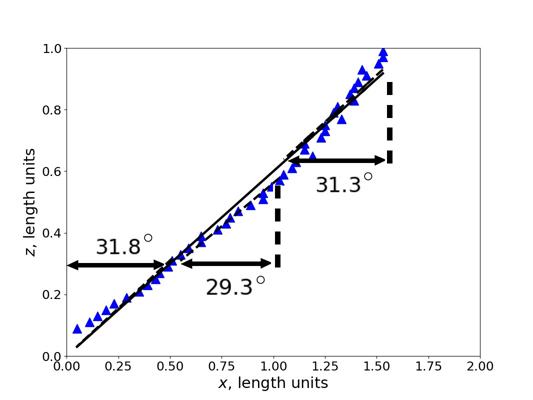}
\end{minipage}%
\begin{minipage}[t]{0.33\textwidth}%
\includegraphics[scale=0.3]{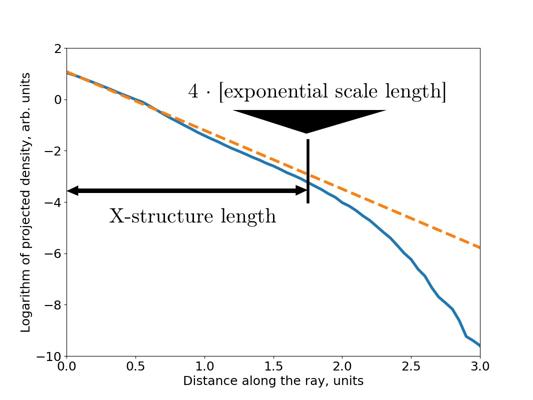}
\end{minipage}%
\caption{Measurement of the X-structure parameters. Data for the model with $M_\mathrm{h}=1.5$ at the time moment $t=7.5$~Gyr (570 time units). \textit{Left}: the distribution of the projected density $\rho_{xz}$ (along the minor axis of the bar) versus the abscissa $x$ in several $z$-cuts. Each solid line corresponds to each $z$-cut, dashed lines correspond to the filtered density (see text for details), triangles mark locations of maximums of the filtered density which correspond to the location of the of X-structure. \textit{Middle}: $(x, z)$ coordinates of the averaged ray of the X-structure (triangles). 
The solid line is the best fit approximation. The arctangent of its angular coefficient is the opening angle of the X-structure. \textcolor{black}{Dashed lines correspond to the best fit lines for different regions: inner (from $0$ to $0.5$), middle (from $0.5$ to $1$), and outer (from $1$ to the last density peak). Labels denote the values of the opening angle for each area.}
\textit{Right}: The logarithm of the projected density measured along the ray of the X-structure. \textcolor{black}{The dashed line corresponds to the best fit exponential curve. The arrow indicates the measured length of the ray (see text for details).}
}
\label{fig:x_params}
\end{figure*}

\begin{figure*}
\begin{minipage}[t]{0.45\textwidth}%
\includegraphics[scale=0.25]{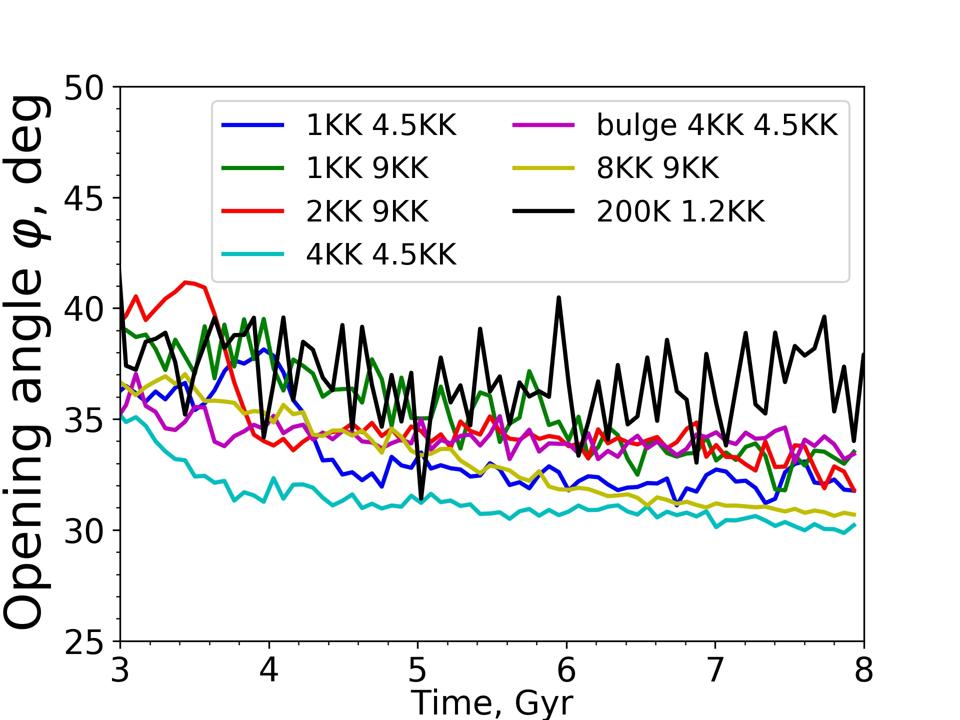}
\end{minipage}
\hfill
\begin{minipage}[t]{0.45\textwidth}%
\includegraphics[scale=0.25]{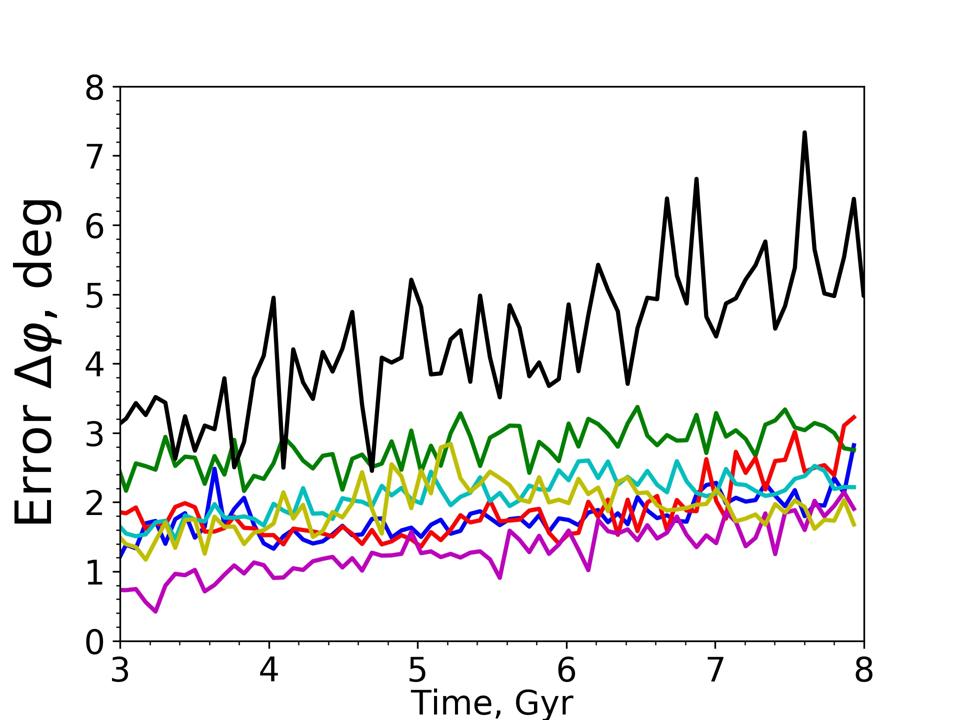}
\end{minipage}
\caption{The evolution of the opening angle of X-structures (left) and its error of measurement (right) in models with $M_\mathrm{h}=1.5$ and different number of particles in the disc and the halo.}
\label{fig:phi_num}
\end{figure*}
\par
To estimate parameters of X-structures, we performed the analysis of modelled galaxies similar to that of observed ones \citep{Savchenko_etal2017} with a few differences. Since the simulated galaxies have a much better spatial resolution than the images of observed galaxies we can avoid the time-consuming procedure of galaxy decomposition and work with the simulated snapshots as a whole because the X-structure rays are clearly detected. Such a shortcut allows us to perform the analysis of X-structures for many moments of time, in the present work each $5$ time units or $66$ Myr, and trace the evolution of parameters of X-structures for each constructed model.
\par
To obtain the value of the opening angle we calculated the projection density of the modelled galaxy along the minor axis of the bar ($x-z$ projection). For this purpose we used a square grid defined by Eq.~\eqref{eq:grid} with $x_{\mathrm{min}}=-3$, $x_\mathrm{max}=3$ (instead $y_{\mathrm{min}}$ and $y_\mathrm{max}$) and $z_\mathrm{min}=-1$, $z_\mathrm{max}=1$. Next, we averaged obtained values of projection density over four quadrants. 
\par 
We defined the location of a ray of the X-structure at a height $z_i$ as the abscissa of the density peak in each layer of constant height $z_i$. To find this peak, we cut one layer with height equal to $z_i$ and thickness equal to $\Delta z$ and found the distance $x_{i, \mathrm{max}}$ at which the density has its maximum in this layer. Since the projection density $\rho_{xz}(x, z)$ has some discontinuity (due to the discrete grid) we smoothed it by Savitzky-Golay filter \citep{Savitzky_Golay1964} and found the extremum of a smoothed curve (see Fig.~\ref{fig:x_params}, left plot). Our analysis shows that the unambiguous density peak can be identified only if one considers the dependence of projection density on the distance from the centre in the layer of constant height. The inverse scheme where one analyses the dependence of the density on the height in the layer of constant distance does not work. There is only one density peak in the layer of constant distance from the centre and this is the density peak at $z_i = 0$ in the disc plane. Secondary small peaks are indistinct and totally smeared by the big ones. 
\par
Fig.~\ref{fig:x_params}~(middle panel) demonstrates the typical dependence of the height on the distance from the centre for the averaged ray of the X-structure for the benchmark model with $M_\mathrm{h}=1.5$ at the time moment $t=570$ (7.5 Gyrs).
\textcolor{black}{
We determined the opening angles of rays, based on the assumption that they come out of  the centre. Inner and outer area density peaks ($x<0.5$ and $x>1.0$, respectively) lay slightly above the best fit line, while middle area peaks ($0.5<x<1.0$) are slightly below it. In observations, some areas can be inaccessible because of dust or low signal-noise ratio, so it is useful to check how values of the opening angle will change if we exclude associated density peaks. As shown in the figure, the resulting opening angle changes by 1$^\circ$--2$^\circ$.
}
\par
\textcolor{black}{\citet{Bureau_etal2006} used the procedure of unsharp-masking to images of edge-on galaxies and noted that X-structures can be divided into two groups, centred and off centred. Visual inspection of our models shows that they do not demonstrate off centred morphology, however for stronger confidence the same unsharp-masking procedure should be applied to our snapshots. We are not completely sure that we will get the same morphology of the rays. Moreover, in Sec.~\ref{sec:sec_observations}, a curious example is given --- different procedures for determining the opening angle give angles that differ by $20^{\circ}$.
}
\par The flatness of the X-structure or the tangent of the opening angle is obtained as the angular coefficient of the best fit line:
\begin{equation}
z = \tan \varphi \cdot x \,.
\end{equation}
We took the standard deviation $\Delta \varphi = \arctan\sqrt{\sum_i (z_i-\tan \varphi \cdot x_i)^2/K}$, where $K$ is the number of fitted data points, as the error of measuring of the opening angle. Generally, the error is about  $1^{\circ}-2^{\circ}$ but can greatly exceed these values in models with low spatial resolution (see Fig.~\ref{fig:phi_num}, right panel).
\par

\par
\textcolor{black}{
For real galaxies the length of X-structure rays is defined as the distance measured along the ray from the centre to the specific isophote.  Magnitude of this isophote is determined from the background intensity and its variation \citep{Savchenko_etal2017}. Such a threshold does not exist in modelled galaxies. 
To measure the length of the X-structure in a comparable way and regardless of the number of particles, it is necessary to use some reasonable assumptions to limit the length of rays or to produce a synthetic background.
Fig.~\ref{fig:x_params} (right panel) gives a hint how this can be done for our models. As shown in the figure, the projected
density along the ray of the X-structure follows an exponential law. Thus, we define the length of the X-structure as $L=4R_\mathrm{X}$, where $R_\mathrm{X}$ is the exponential density scale of the X-structure along the ray.
} 

\subsection{Time evolution of the opening angle} \label{sec:phi_evol}
The next question considered in this section is how the relaxation, which clearly affects the structure of the disc in the vertical direction, can influence the X-structure.
To determine its influence, we consider what differences in the opening angle can arise in the same models that differ only by the number of the disc and halo particles. The evolution of the opening angle in the model with $M_\mathrm{h} = 1.5$ with different sets of particles is shown in Fig.~\ref{fig:phi_num} (right plot). The first good thing to note is that the behaviour of all curves generally coincides. All newly born X-structures have a large opening angle ($35^\circ-40^\circ$). As the time passes, the opening angle decreases with time. At $t = 8$~Gyr the value of measured opening angle falls between 30$^\circ-32^\circ$. The model with the smallest number of particles in the disc, $N_\mathrm{d} = 2 \cdot 10^5$, is an exception and, for this model, the dependence of the opening angle on time is quite peculiar. As the number of particles is small in this model, the number of knots for fitting the ray of the X-structure is small too. It results in an increase of uncertainties in measuring the position of the X-structure ray which leads to a saw-toothed shape of the dependence of the opening angle on time and large errors (up to $7^{\circ}$, Fig.~\ref{fig:phi_num}, right panel). Aside from this model with low spatial resolution, measured values of the opening angle  coincide within the error of the measurement ($\pm 2^{\circ}$) for models with $N_\mathrm{d}>10^6$.

\section{Dependence on physical parameters} 

\begin{figure*}
\includegraphics[scale=0.1]{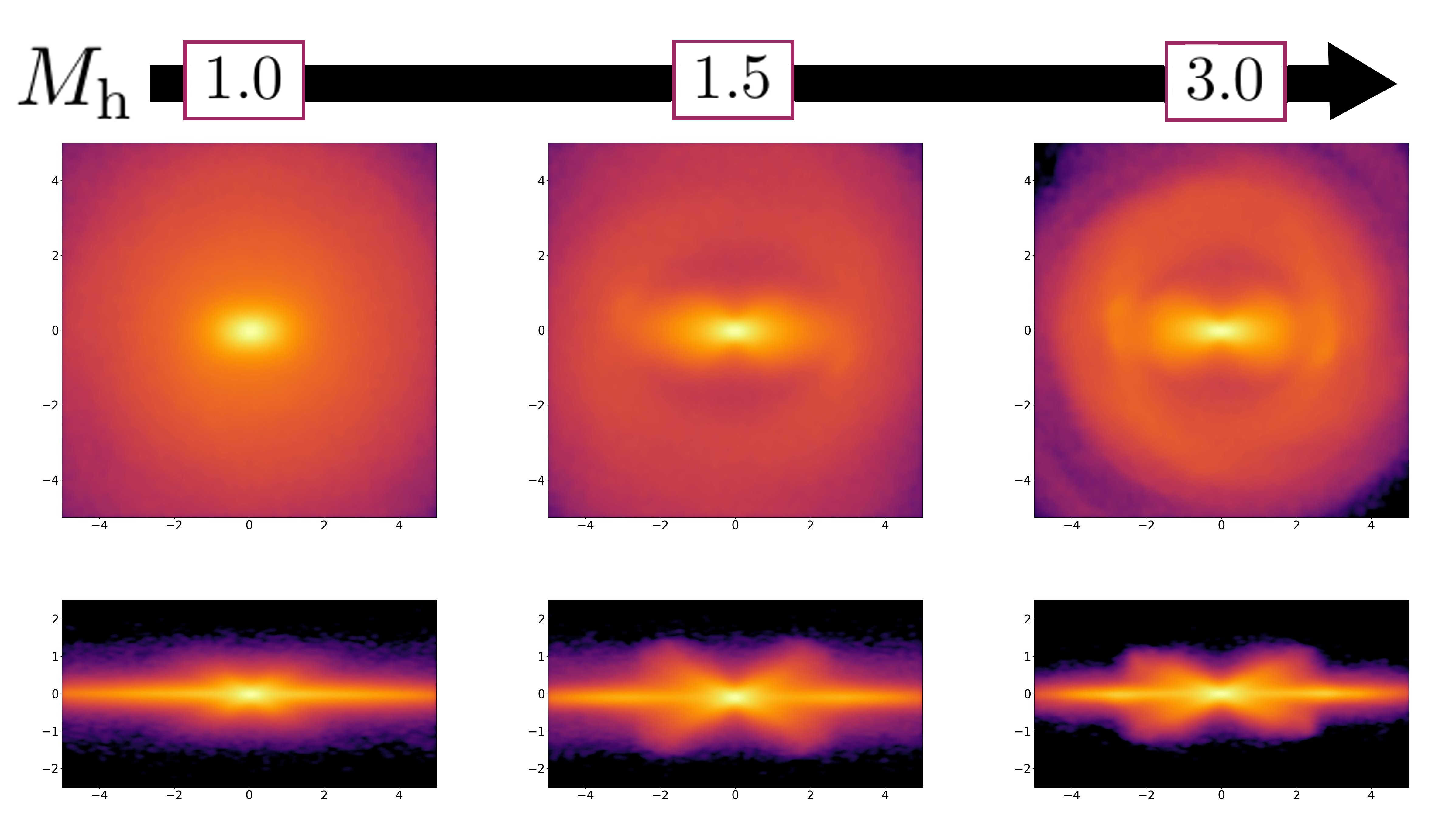}
\caption{The in-plane view (top row) and the edge-on view (bottom row) of models with different halo masses at $T=7.9$ Gyr 
\textcolor{black}{shown in simulations length units}. From $M_\mathrm{h}=1.0$ to $M_\mathrm{h}=1.5$, an X-structure becomes substantially pronounced and more flattened while an in-plane bar got a peanut shape.}
\label{fig:im_halo}
\end{figure*}

\begin{figure*}
\includegraphics[scale=0.1]{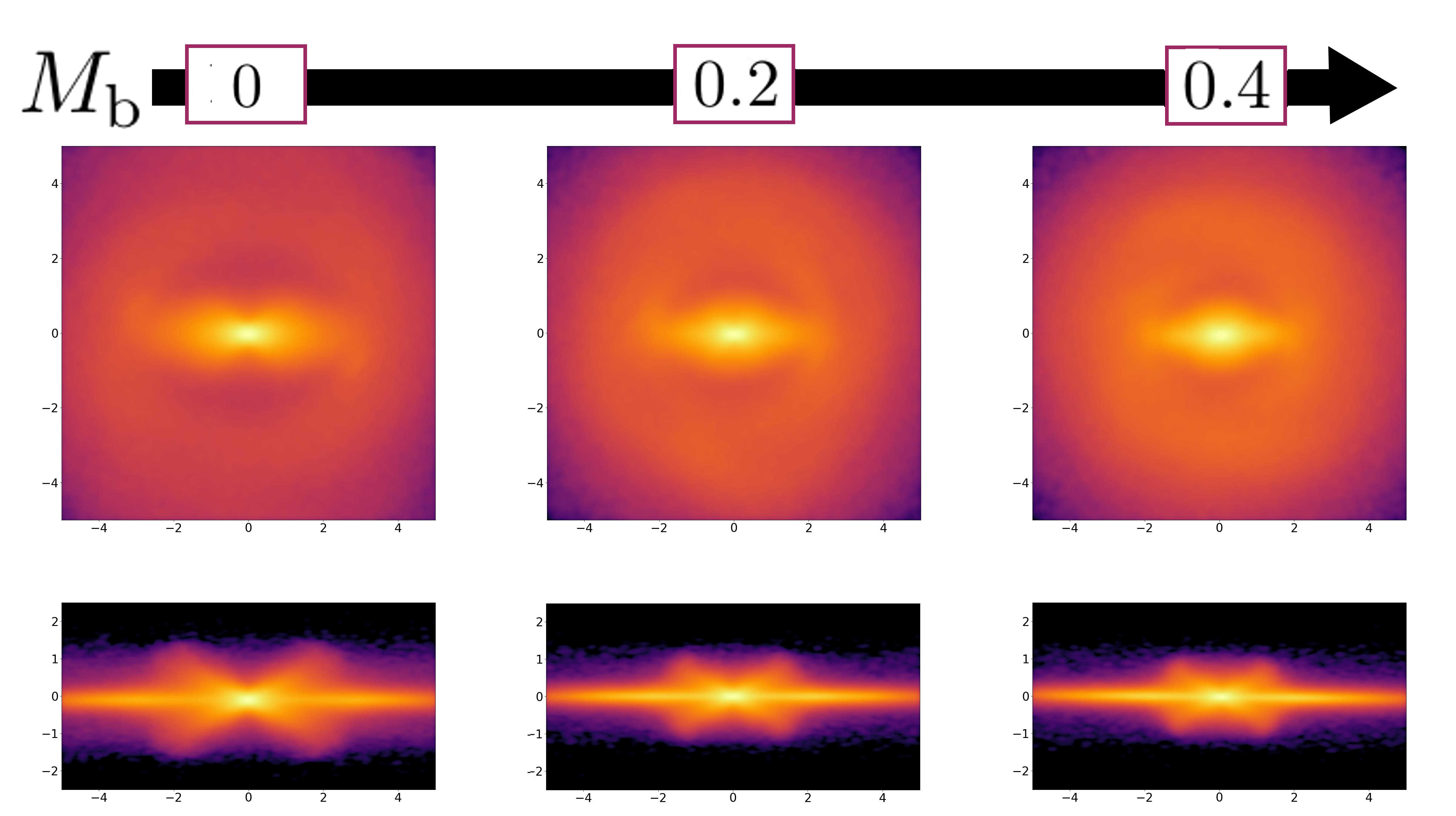}
\caption{The same as in Fig.~\ref{fig:im_halo} for models with different bulge contributions. With an increase of bulge mass, X-structures become smaller and an in-plane bar transforms into a barlens.}
\label{fig:im_bulge}
\end{figure*}

\begin{figure}
    \centering
    \includegraphics[scale=0.72]{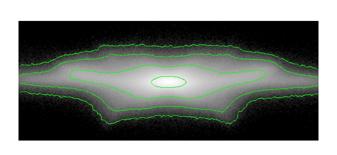}
    \caption{Example of secondary rays for the model with $M_\mathrm{h}=1.5$, $Z_\mathrm{d}=0.05$, $Q=2.0$ at $T=7.6$ Gyr.}
    \label{fig:secondary_x}
\end{figure}

\begin{figure*}
\includegraphics[scale=0.1]{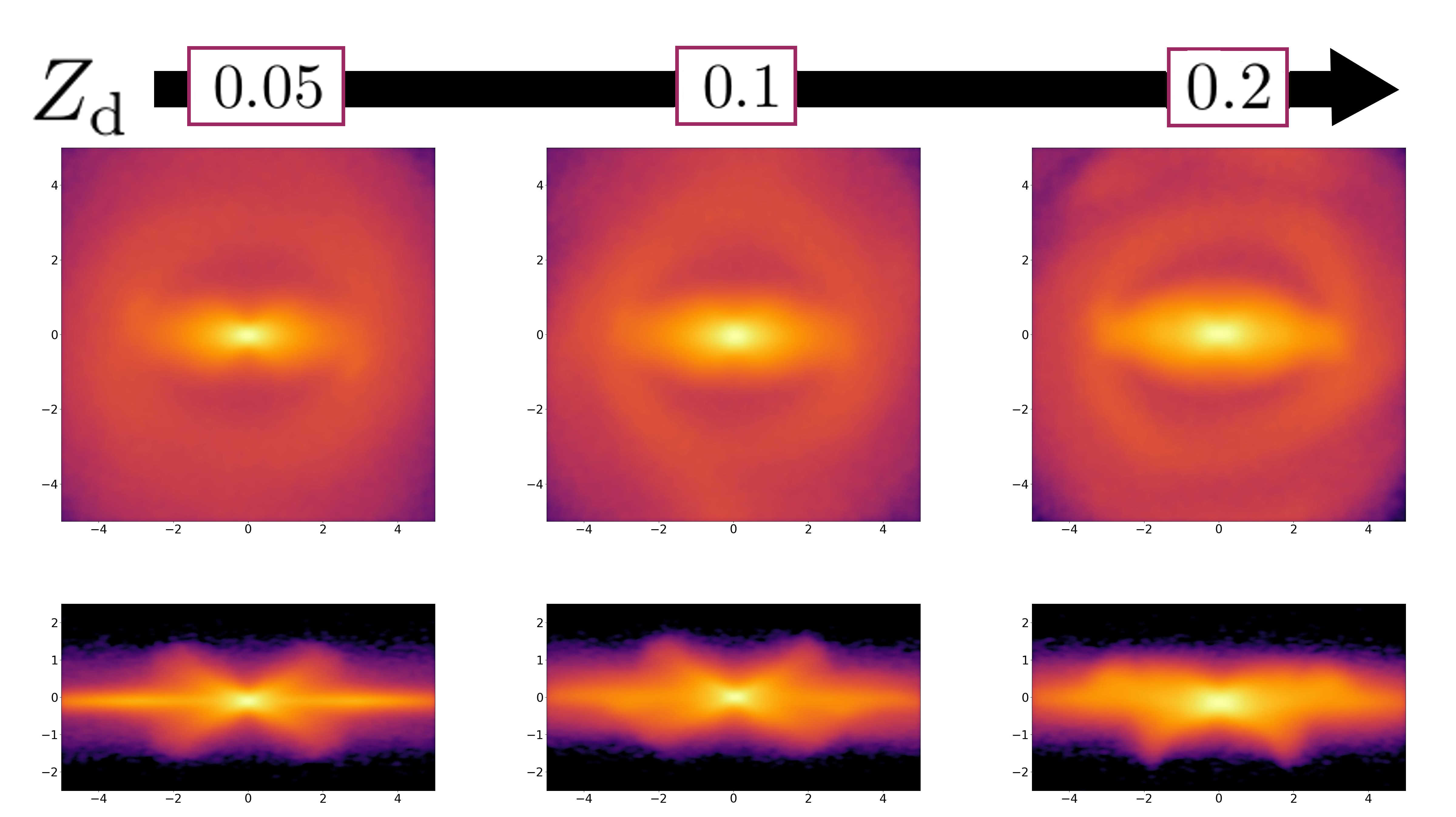}
\caption{The in-plane view (top row) and the edge-on view (bottom row) of models with different values of the initial thickness of a stellar disc at $T=7.9$ Gyr. From $Z_\mathrm{h}=0.05$ to $Z_\mathrm{d}=0.1$, X-structures become slightly steeper, more asymmetric, an in-plane bar gets the barlens morphology.}
\label{fig:im_thickness}
\end{figure*}

\begin{figure*}
\includegraphics[scale=0.1]{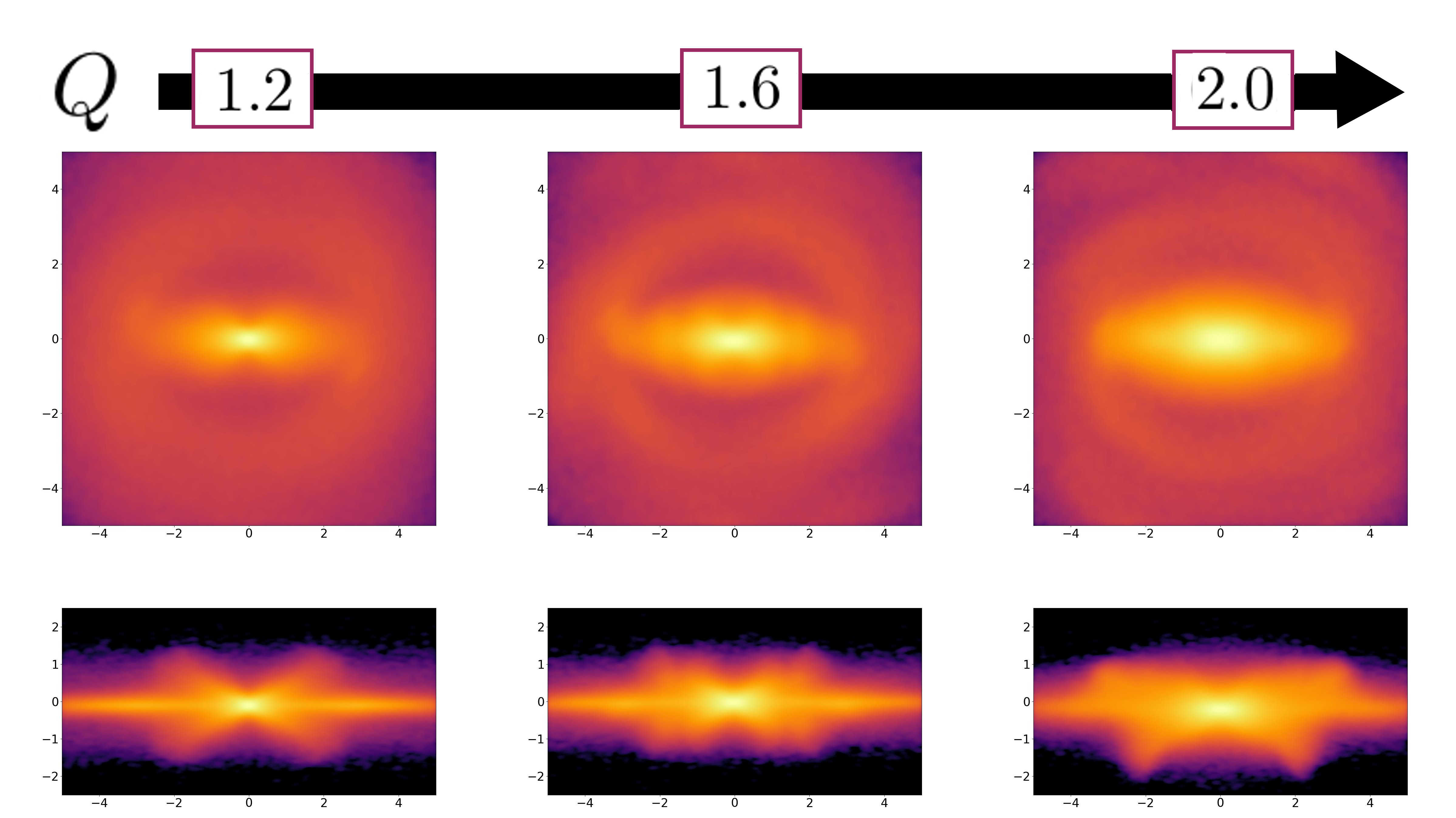}
\caption{The same as in Fig.~\ref{fig:im_thickness} for models with different values of the Toomre parameter. From $Q=1.2$ to $Q=1.6$, an X-structure transforms into a symmetric \textit{double} X-structure. For $Q=2.0$, secondary buckling is long-lasting and a bar has not regained its symmetry. An increase of the Toomre parameter leads to barlens morphology.}
\label{fig:im_toomre}
\end{figure*}

\begin{figure*}
\includegraphics[scale=0.1]{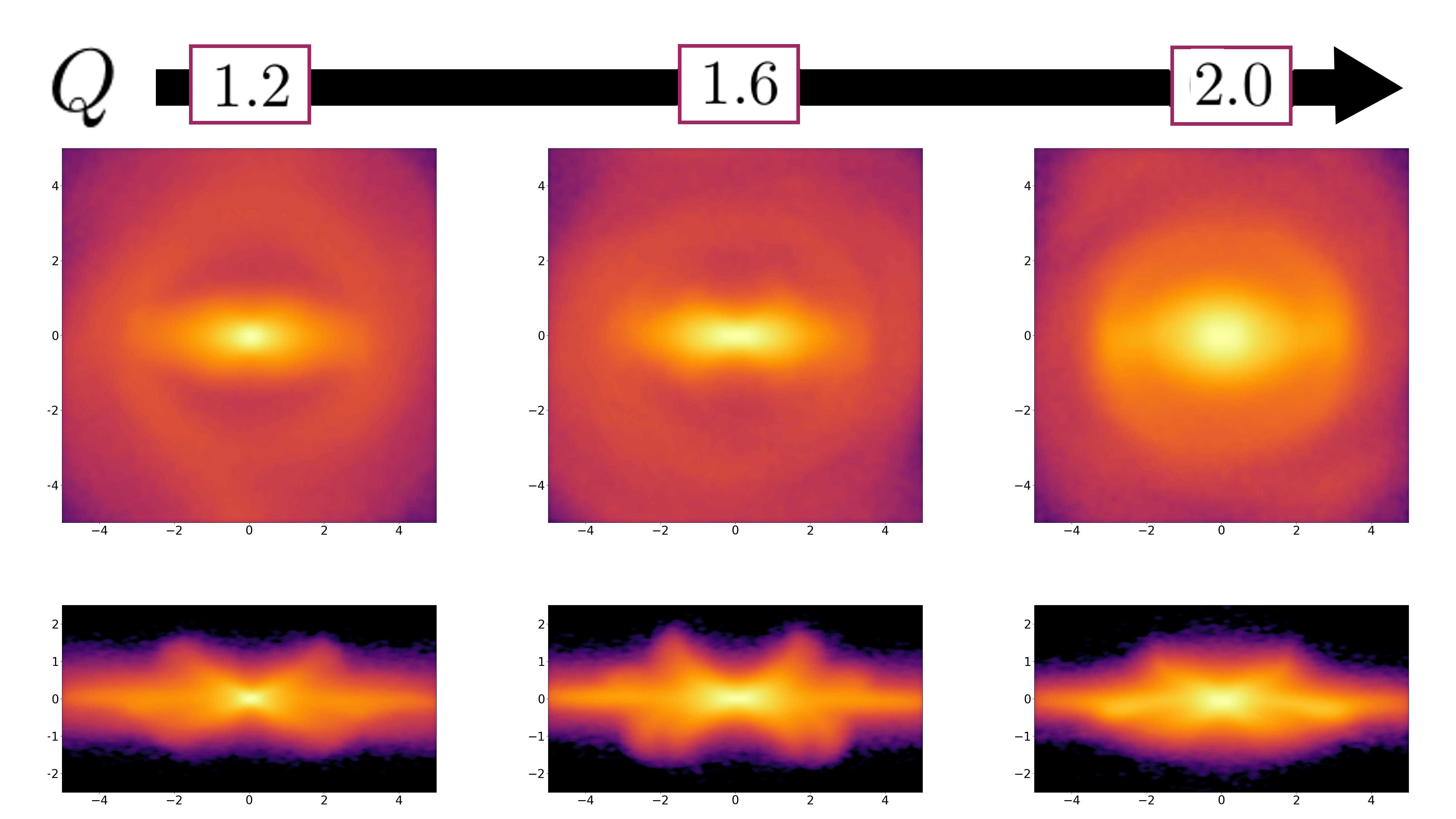}
\caption{The same as in Fig.~\ref{fig:im_toomre} for models with different values of the Toomre parameter, but for initially thick disc, $Z_\mathrm{d}=0.1$. As it was in thin discs, from $Q=1.2$ to $Q=1.6$, an X-structure transforms into a double X-structure but in this case outer X-structures are yet strongly buckled. Model with $Q=2.0$ demonstrate a strongest bar-lens morphology of observed in out set of models. Also note that a bar surrounding ring become a very prominent too in this model.}
\label{fig:im_toomre2}
\end{figure*}

\begin{figure}
\begin{minipage}[t]{0.45\textwidth}%
\includegraphics[scale=0.28]{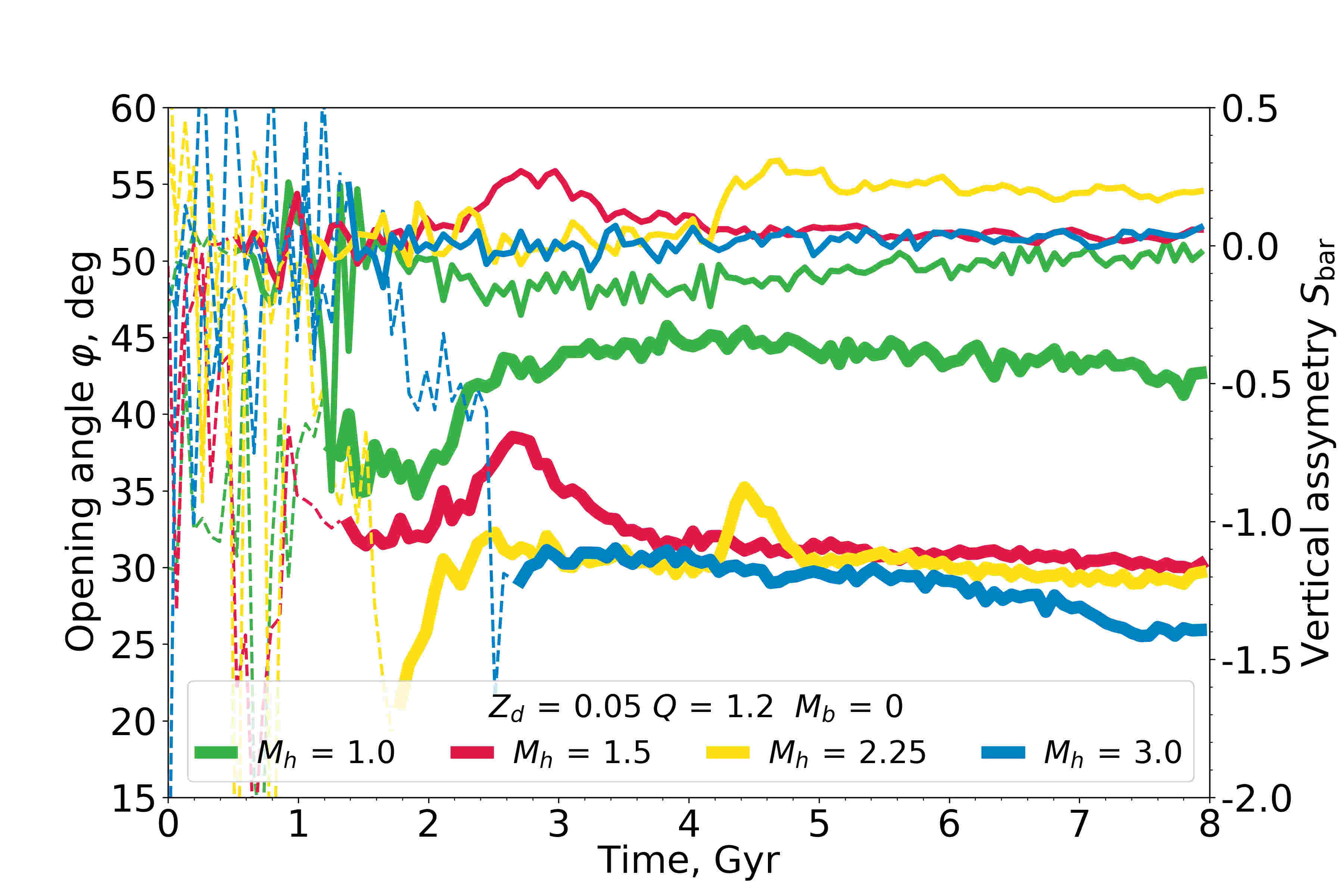}
\end{minipage}
\caption{The dependence of opening angles (left axis, thick lines) and the vertical asymmetry parameter (right axis, thin lines) on time. For opening angles, dashed lines denote the period prior to the X-structure formation. In a similar way for an asymmetry parameter, dashed lines denote the period when a bar harmonic is relatively weak, ($A_2/A_0<0.05$). }
\label{fig:phi_halo}
\end{figure}

\begin{figure}
\begin{minipage}[t]{0.45\textwidth}%
\includegraphics[scale=0.28]{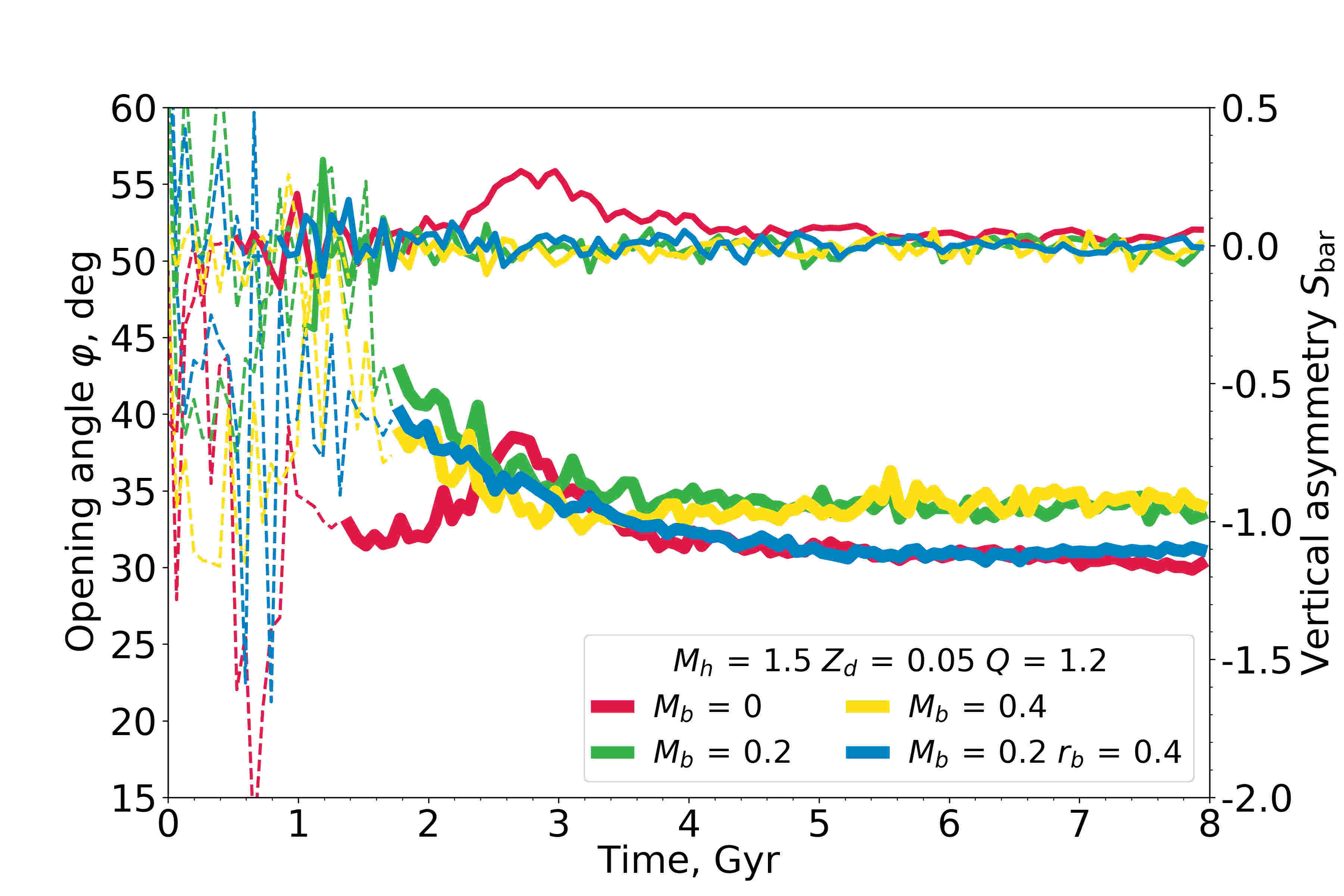}
\end{minipage}
\caption{The same as in Fig.~\ref{fig:phi_halo} for models with different bulges. }
\label{fig:phi_bulge}
\end{figure}

\begin{figure}
\begin{minipage}[t]{0.45\textwidth}%
\includegraphics[scale=0.28]{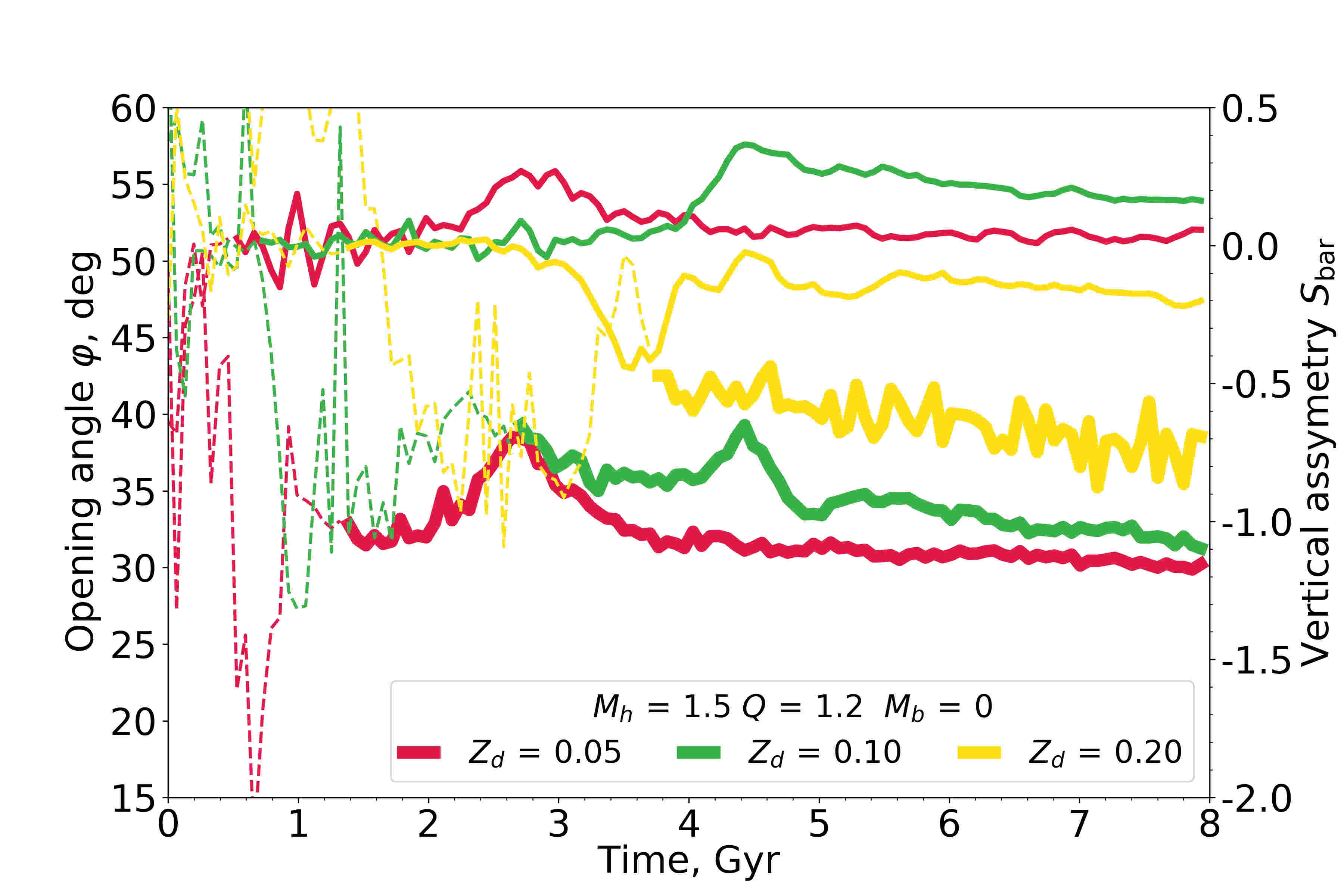}
\end{minipage}
\caption{The same as in Fig.~\ref{fig:phi_halo} for models with different values of disc thickness. An increase of opening angles is connected with an increase of the disc thickness.}
\label{fig:phi_z}
\end{figure}

\begin{figure}
\begin{minipage}[t]{0.45\textwidth}%
\includegraphics[scale=0.28]{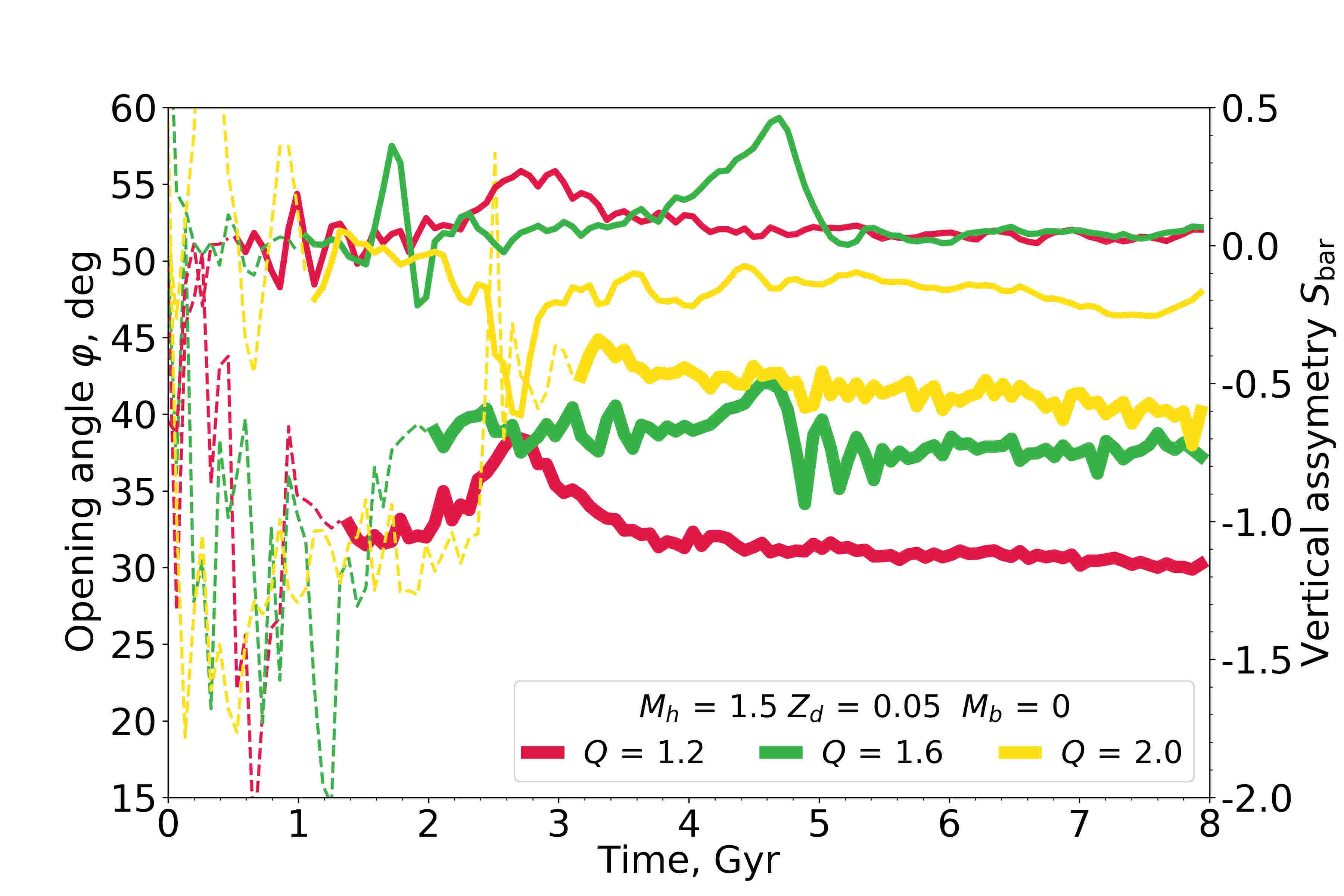}
\end{minipage}
\caption{The same as in Fig.~\ref{fig:phi_halo} for models with different values of the Toomre parameter. Opening angles are increasing with an increase of the Toomre parameter.}
\label{fig:phi_toomre}
\end{figure}

\begin{figure}
\begin{minipage}[t]{0.45\textwidth}%
\includegraphics[scale=0.28]{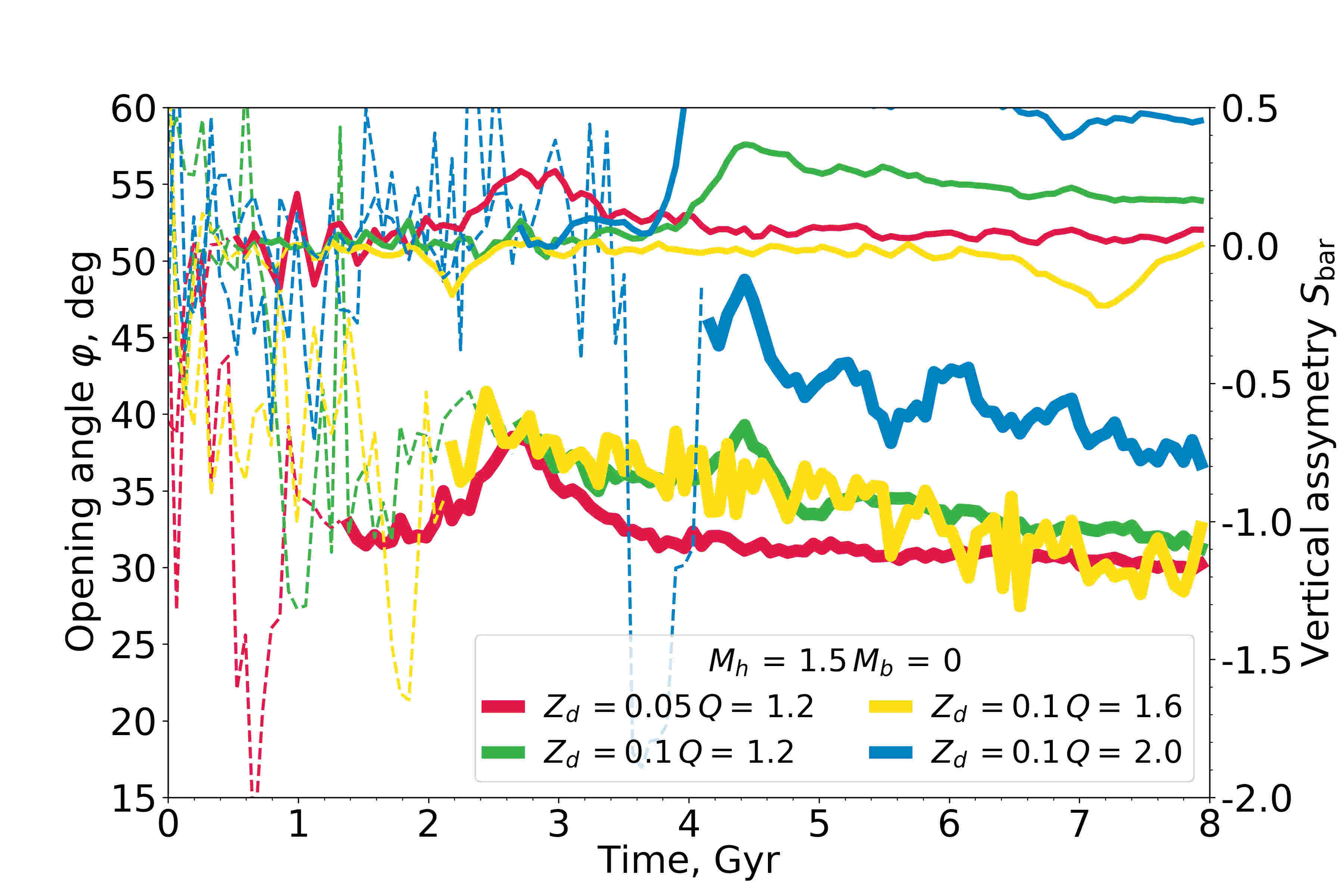}
\end{minipage}
\caption{The same as in Fig.~\ref{fig:phi_halo} for models with different values of the Toomre parameter and initially thick disc, $Z_\mathrm{d}=0.1$. Note a very strong buckling in the model with $Q=2.0$.}
\label{fig:phi_toomre2}
\end{figure}
\par
All bars in models with different physical parameters from Table~\ref{tab:models_mu} end up with X-shaped structures if the view is edge-on. Though bars share such clear similarities, in-plane bars and X-structures viewed edge-on significantly differ both in morphology and evolutionary paths due to initial difference in physical parameters of galaxy models. The first and the most eye-catching peculiarity is secondary long-lasting strong buckling. If it arises, there are two possible scenarios. First, if buckling is not strong enough then an X-structure bends to the one side of the disc. After about $1-2$ Gyrs the whole system returns to a symmetric not perturbed state and after that, there is a slow and gradual growth of a bar and an X-structure. In hot models (with $Z_\mathrm{d}=0.1;0.2$ or $Q=1.6;2.0$) buckling is so strong that additional bright rays start to stand out from the disc plane (Fig.~\ref{fig:im_thickness} and Fig.~\ref{fig:im_toomre}). Visually, such rays contain far more stars than the primary X-structures rays. For these models, a buckled state lasts at least $4$ Gyrs (this time limit arises because such models are buckled till the last snapshot of simulations, see Fig.~\ref{fig:im_thickness} and Fig.~\ref{fig:im_toomre}). With a characteristic time about $1$ Gyr, secondary rays form \textit{secondary} X-structure which is far brighter and thicker than the primary one. \textcolor{black}{Such secondary rays are especially noticeable in  the grey-scale intensity map with density contours (see Fig.~\ref{fig:secondary_x}).}
In this work, we manually exclude such secondary X-structures from our analysis if they appear. As they pollute a distribution of density maximums which describes the primary X-structure only after some certain distance ($R > 1.0 R_\mathrm{d}$), we simply constraint boundaries of the grid on which the density profile is calculated. 
\par 
\textcolor{black}{
Another interesting feature is associated with hot and thick models. Model with $Q=1.6$ and $Z_\mathrm{d}=0.1$ produces a slightly off-centred X-structure, see Fig.~\ref{fig:im_toomre2}. The trick is that though visually it is indeed off-centred, an accurate study of density peaks distribution indicates that they still come out of the centre of the disc, but have a larger curvature of the density peak curve in the central regions ($0.5<x<1.0$) than it is observed in a fiducial model with a cool disc (see Fig.~\ref{fig:x_params}, middle panel). Therefore we can determine the opening angle associated with such an X-structure in the same way as for other models.
}
\par
We use a routine described in the previous section to measure opening angles of all model with different physical parameters from Table~\ref{tab:models_mu}. To accurately trace the influence of each varied parameter, we consider each set of models associated with it individually.
\par
\textcolor{black}{We measure the buckling strength by the parameter of vertical asymmetry, $S_\mathrm{bar} = (A_2(z>0)-A_2(z<0))/A_2$}, where $A_2(z>0)$ is the amplitude of a bar above the disc plane ($z>0$), whereas $A_2(z<0)$ is the amplitude below the disc plane ($z<0$). The simple physical meaning of this parameter is how much stronger/weaker a bar harmonic on one side of the disc plane than the harmonic on the other side in terms of the bar total amplitude. 

\subsection{Dependence on halo mass}
\label{sec:phi_halo}
The end-state structure of bars for different halo masses (in-plane and side-on views) is shown in Fig.~\ref{fig:im_halo}. The evolution of opening angles and a vartical asymmetry parameter in four models with the different halo contribution are shown in Fig.~\ref{fig:phi_halo}. All bars are undergoing a short phase of buckling ($\approx 0.5$ Gyr) after which a small X-structure appears\footnote{Our procedure of measuring of opening angles carries a physical meaning only when density maximums form a straight line like those in Fig.~\ref{fig:x_params}. From the start of simulations, density maximums are distributed randomly. At some point in time, they begin to line up and one can roughly define this moment as the time of the birth of an X-structure. We determine this moment of time from visual checking of the evolution of density maximums in each model.}. After that, a newly born X-structure grows in size and flattens. A model with the lightest halo, $M_\mathrm{h}=1.0$, follows another scenario. At first, an opening angle is about $36^\circ$. After $2$ Gyr, it suddenly increases to about $42^\circ$ and remains at this level till $T=7.9$ Gyr. Such an unusual behaviour probably associated with dissolving of a first bar at $1.5$~Gyr which occurs in this model and formation and buckling of the next bar which is much smaller and has a barlens morphology (Fig.~\ref{fig:im_halo}, upper left plot). Nevertheless, an X-structure in this model turns out to have the greatest value of the opening angle, $\varphi=40^{\circ}$ at all times in comparison with other models. In contrary, the smallest observed angle in our simulations, $\varphi = 26^\circ$ emerges in a model with the heaviest halo $M_\mathrm{h}=3$. Besides, even a simple visual comparison of X-structure view in different models indeed reveals that model with the heaviest halo has a more flattened X-structure (see Fig.~\ref{fig:im_halo}, bottom row). Two models with intermediate halos have opening angles with values between previous two, $\varphi \approx 30^\circ$. 
\par
As can be seen from the evolution of asymmetry parameter in Fig.~\ref{fig:phi_halo} (upper curves), all models except one with the heaviest halo are undergoing the long-lasting phase of buckling. For the model with $M_\mathrm{h}=2.25$, buckling lasts until the end of simulations. For  models with $M_\mathrm{h}=1.0$ and $M_\mathrm{h}=1.5$, it lasts for about $2$ Gyrs and after that, a bar returns to the unperturbed state. A direct comparison of values of opening angles and asymmetry parameter reveals that as soon as buckling arises the values of the opening angle bounces up to about $5^{\circ}$. Though secondary buckling presents, an overall morphology of X-structures is similar for all models with different halo mass (with the exception of the model with $M_\mathrm{h}=1.0$), which is not nearly the case for models with different values of thickness and Toomre parameter. (see below).

\subsection{Dependence on central concentration}
As can be seen from Fig.~\ref{fig:im_bulge} (bottom row), models with substantial bulge contribution have smaller and tighter X-structures in comparison with a bulgeless model. Such X-structures have greater opening angles, by about $4^\circ-5^\circ$ (see Fig.~\ref{fig:phi_bulge}, bottom curves). There are no any traces of secondary buckling which is typical for almost all other models (Fig.~\ref{fig:phi_bulge}, top curves), though there is small rapid initial buckling which takes place before the X-structure formation. As X-structures appears, they lower their opening angle, from $40^{\circ}$ to $34^{\circ}$.  After $4$ Gyrs, values of opening angle seem to saturate and they almost do no change till the end of simulations. Note that the model with a light and diffuse bulge ($M=0.2, r_\mathrm{b}=0.4$) ends up with almost the same value of the opening angle as the bulgeless model. Though it is not the primary aim of the present work, we also note that the bulge presence changes an in-plane bar morphology from peanut into barlens (Fig.~\ref{fig:im_bulge}, top row), which is consistent with results of the recent work by \cite{Salo_Laurikainen2017_v2}.

\subsection{Dependence on initial disc thickness}
Visually X-structures in models with high initial disc thickness ($Z_\mathrm{d}=0.1$ and $Z_\mathrm{d}=0.2$) are quite different from those in thin models with a bulge or a massive halo. The most outstanding feature of such models is secondary (additional to primary) X-structures which appear to be a consequence of violent secondary buckling (Fig.~\ref{fig:im_thickness}, bottom row). Unfortunately, our procedure of measurement of the opening angle cannot be applied to secondary structures because they do not give sufficient contribution to the density distribution to distinguish secondary density peaks (at least near the disc plane). In Fig.~\ref{fig:phi_z} we show obtained values of opening angles of the primary (inner) X-structure and an asymmetry parameter of a bar. An increase of the disc thickness by a factor of two from $Z_\mathrm{d}=0.05$ to $Z_\mathrm{d}=0.1$ leads to a slight increase of the opening angle, by about $3^\circ$, while for the model with $Z_\mathrm{d}=0.2$, the overall increase of the opening angle is considerably greater, such that at $T=7.9$~Gyr it is approximately equal to 40$^\circ$. These models are also subjects to the violent buckling instability that lasts at least for $4$ Gyrs (Fig.~\ref{fig:phi_z}, upper curves). As for an in-plane morphology,  a bar evolves in a barlens with an increase of disc thickness too (Fig.~\ref{fig:im_thickness}, top row).

\subsection{Dependence on the Toomre parameter}
Our sample of models with different values of the Toomre parameter contains two subsamples of models with initially thin ($Z_\mathrm{d}=0.05$) and initially thick ($Z_\mathrm{d}=0.10$) discs. In both types of models, an increase of the Toomre parameter leads to changes of bar vertical and in-plane morphology similar to those arising with an increase of disc thickness. From $Q=1.2$ to $Q=1.6$, a secondary buckling and secondary X-structures appear (Fig.~\ref{fig:im_toomre} and Fig.~\ref{fig:im_toomre2}, bottom rows). Analogous to the influence of an increase of bulge mass and disc thickness, an increase of $Q$ leads to the barlens bar morphology seen in-plane (Fig.~\ref{fig:im_toomre} and Fig.~\ref{fig:im_toomre2}, top rows).
\par
\textit{Thin disc:} Opening angles increase with an increase of the Toomre parameter, starting from 36$^\circ$ in the model with $Q=1.6$ to 40$^\circ$ in the model with $Q=2.0$ (Fig.~\ref{fig:phi_toomre}, bottom curves). The model with $Q=1.6$ is quite interesting because though vigorous secondary buckling has taken place in this model (Fig.~\ref{fig:phi_toomre}, upper curve), the whole system returns to a symmetric state with a double X-shaped structure (Fig.~\ref{fig:im_toomre}, middle bottom plot), which is not observed in our other models. For $Q=2.0$, the secondary buckling is long-lasting and a bar has not regained its symmetry (Fig.~\ref{fig:phi_toomre}, upper curve).
\par
\textit{Thick disc}:   All  X-structures demonstrate a lopsided  (with  respect  to  the galactic  plane) end-state shape. In contrary to the thin disc, an increase of the Toomre parameter from $Q=1.2$ to $Q=1.6$ almost does not change an opening angle of the inner X-structure (see Fig.~\ref{fig:phi_toomre2}). Also, secondary X-structures appear to be asymmetrical till the end of simulations in both models with $Q=1.6$ and $Q=2.0$. The model with $Q=2.0$ shows smaller angles, $\varphi\approx 36^\circ$, instead of $40^\circ$ observed in its thin counterpart. 
\par
Inconsistencies between dependencies of the opening angle on the Toomre parameter in both thin and thick models clearly shows the formation of X-structures is a complicated and not linear process which details had yet to be understood.
\subsection{X-structures sizes}
\begin{figure*}

\begin{minipage}[t]{0.45\textwidth}%
\includegraphics[scale=0.2]{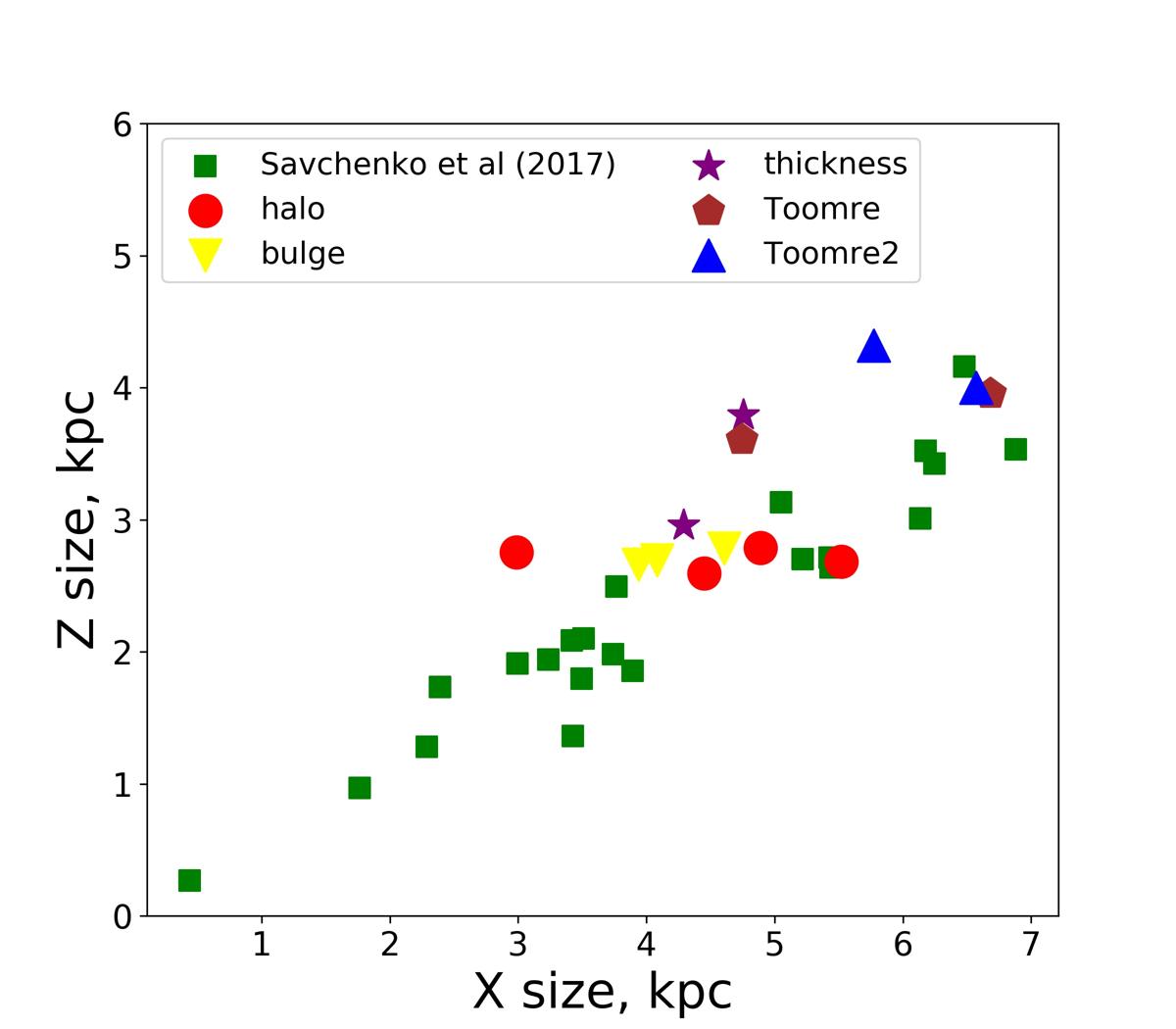}    
\end{minipage}%
\begin{minipage}[t]{0.45\textwidth}%
\includegraphics[scale=0.2]{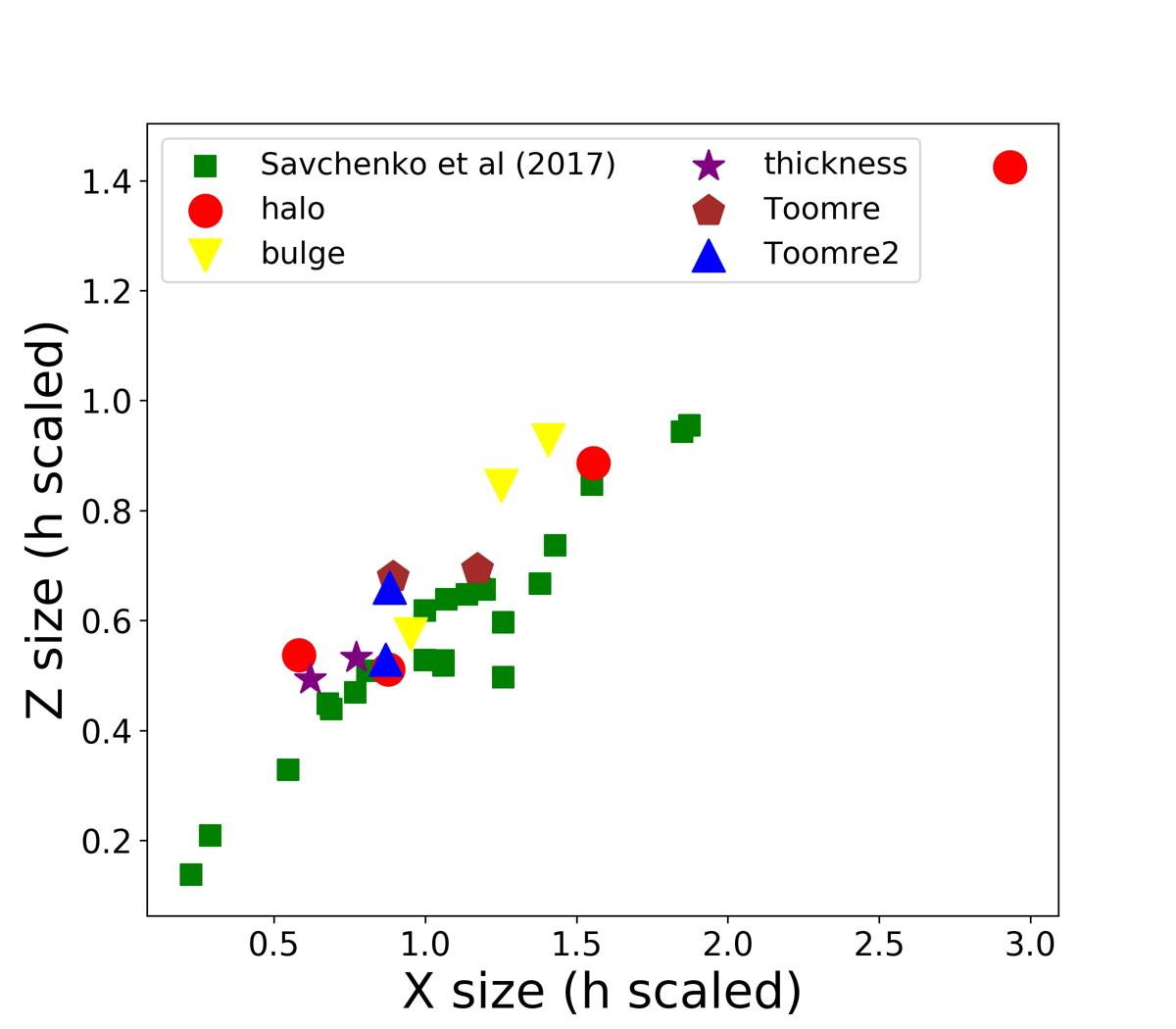}     
\end{minipage}%

\caption{Lengths of projections of X-structures onto $(xy)$ (X size) and $(yz)$ (Z size) planes in kpc (left panel) and in units of an exponential disc scale length (right panel) for $N$-body models considered in the present work and X-structures observed in real galaxies from work by~\citet{Savchenko_etal2017}. Different colours of markers denote different sets of models, see~Table~\ref{tab:models_mu}. "Toomre2" label refers to the set of models with a thick disc $Z_\mathrm{d}=0.1$ and high Toomre paramter values $Q=1.6, \, 2.0$.}
\label{fig:models_xz}

\end{figure*}

From the observational point of view, an X-structure can be described more thoroughly by an additional parameter such as the length of its horizontal and vertical projections, $X = L \cdot \cos\varphi$  and $Z = L \cdot \sin \varphi$, where $L$ is the length of the averaged ray. 
\textcolor{black}{
We display the resulting X-structure sizes at the end of the simulations ($T = 7.9$ Gyr) in two different scales, in absolute values (Fig.~\ref{fig:models_xz}, left panel) and in units normalized to the disc scale length (Fig.~\ref{fig:models_xz}, right panel), since both are commonly used in the literature.
} 
In general, end-state stellar discs in our experiments have complicated structure which does not satisfy a single exponential law. For most models, there are two exponents plus a small hump between them originated from a ring surrounding a bar. We measure the scale of the outer disc and use it as normalisation of the length of X-structures. In this way, we mask an inner area of the disc which is commonly done in observational works where one searches disc parameters in multi-component galaxies. 

\par
\par
In the same figure, we show sizes of real X-structures measured in the work by \cite{Savchenko_etal2017}\footnote{For clarity, we do not use data from work of~\cite{Ciambur_Graham2016} as authors used a different set of parameters to describe X-structures. In particular, they do not use the full length of X-structure rays, which is used in the present work, but the distance to the maximum of an amplitude of perturbation of elliptical isophotes described by sixth Fourier harmonic. Such sizes are smaller than those measured by our method (see Sec.~\ref{sec:measure_x}), but it is unclear how much they differ.}. \textcolor{black}{As can be seen from the figure, real X-structure sizes show greater variance than those of modelled galaxies. A visible offset to greater values of modelled X-structures can be a consequence of different reasons. First of all, observed sizes are limited by the telescope light-gathering power and background intensity. Our procedure of measuring is solely based on X-structure scale length estimation and does not take into account the possible variance of background intensity. Therefore, one can expect a systematic offset of modelled X-structures sizes to greater values. Note the fact that there are no observable X-structures of sizes greater than modelled ones which is definitely in favour of this reasoning. In general, we can conclude that observed and model X-structures have roughly similar sizes. For a more thorough comparison, the same processing procedure should be applied to model and observed galaxies.}    
\par Modelled X-structures sizes demonstrate some dependencies on galaxy parameters. For galaxies with the high mass of a halo ($M_\mathrm{h}=2.25$ and $M_\mathrm{h}=3.0$), we obtain that X-structures are more elongated. The presence of a bulge and a decrease of halo mass shortens X-structures. An increase of disc thickness, as well as an increase of a Toomre parameter, lead to more elongated X-structures (note that we do not include secondary X-structures here which are observed in some models). In terms of disc scale units, sizes obtained from numerical simulations are consistent with observations too.  For most X-structures, an X-size is about 1.0-1.5 of an exponential disc scale length and a Z-size is about 0.5-0.9.
\par
\begin{figure}
\begin{center}
\begin{minipage}[t]{0.45\textwidth}%
\includegraphics[scale=0.2]{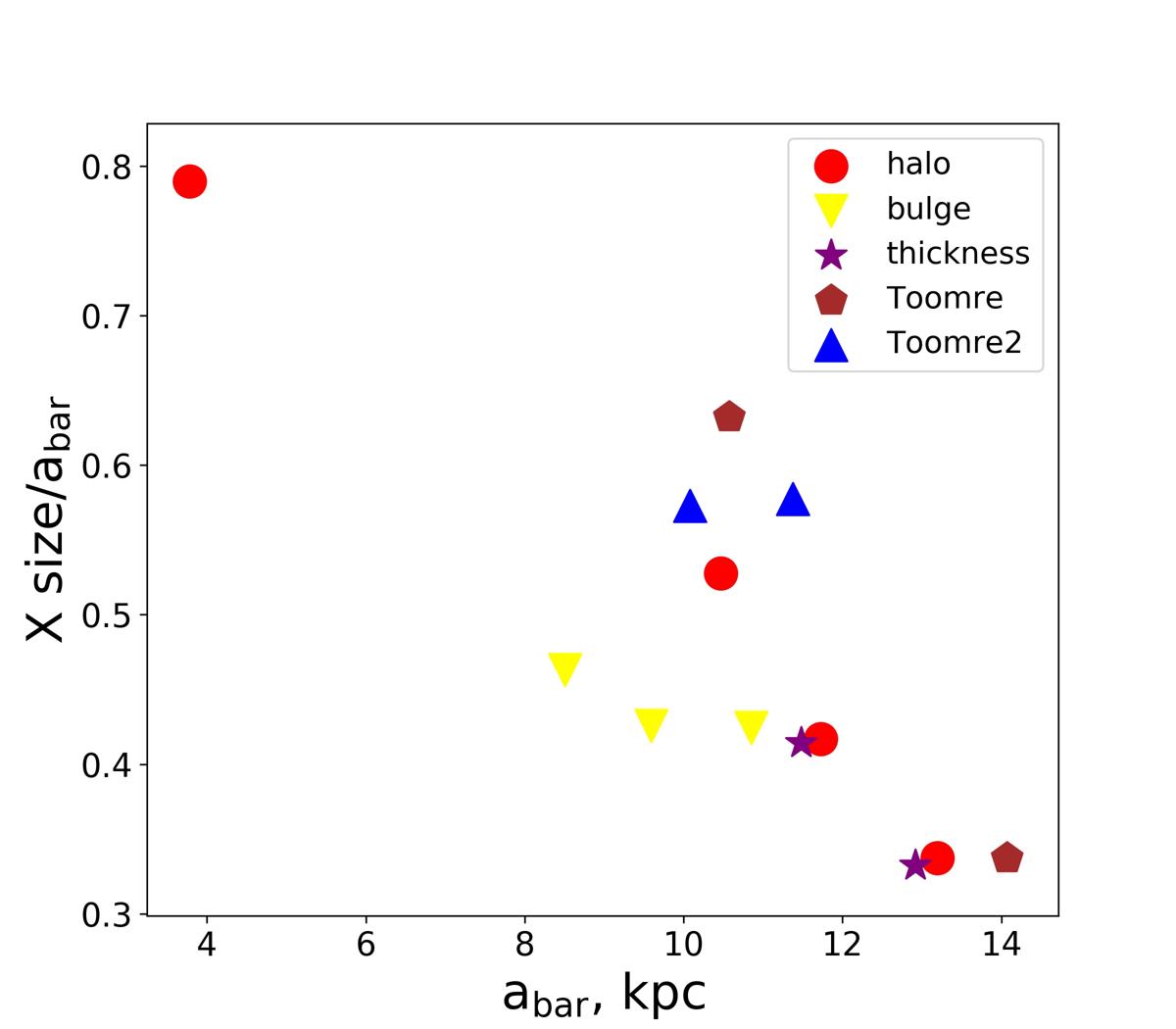}    
\end{minipage}%
\caption{The ratio of modelled X-structure sizes in disc plane (X size) and the half-lengths of a bar major axes versus the half-lengths of a bar major axes.}
\label{fig:models_xz_ratio}
\end{center}
\end{figure}
\par
\textcolor{black}{
In observations, X-structure sizes are commonly compared with the half-length of a bar major axis. The ratio is ranged from 0.3 to 0.8 with a typical value 0.5~\citep{Erwin_Debattista2017}. 
The ratio of the simulated X-structure size projected onto the disc plane (X-size) to the half-length of the bar major axis versus the half-length of the bar major axis is shown in Fig.~\ref{fig:models_xz_ratio}. The X-shape sizes of the models are 0.3 -- 0.8 of the bar length, while most values range from 0.3 to 0.6 which is generally consistent with an observational statistic.
}

\section{Discussion}

\begin{figure}
\begin{center}
\begin{minipage}[t]{0.45\textwidth}%
\includegraphics[scale=0.37]{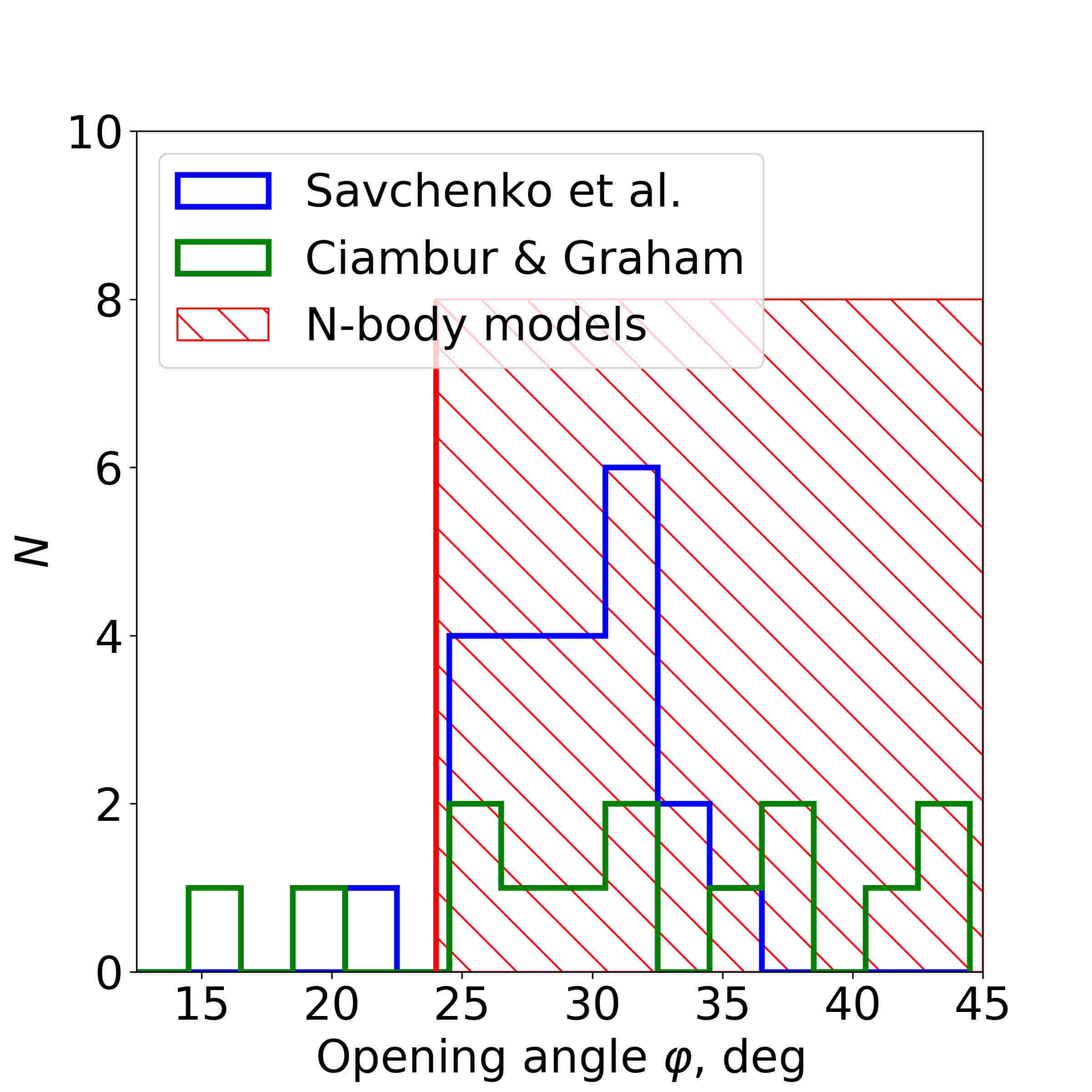}
\end{minipage}
\caption{The histogram  of observed values of opening angles compiled from data by \citet{Ciambur_Graham2016} and \citet{Savchenko_etal2017}. A crosshatched region corresponds to the range of opening angles determined for modelled galaxies.}
\label{fig:phi_hist}
\end{center}
\end{figure}

\begin{figure}
\begin{center}
\begin{minipage}[t]{0.45\textwidth}%
\includegraphics[scale=0.25]{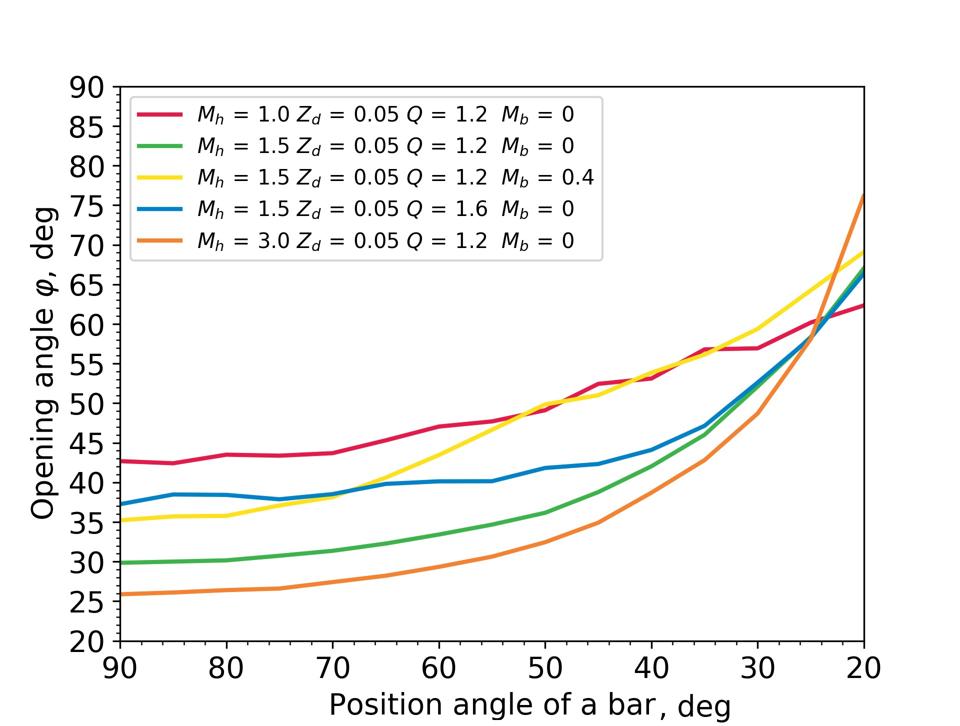}
\end{minipage}%
\caption{The dependence of opening angles on the position angle of a bar major axis for some models. All models, except one with a bulge, show similar initial trend, for a decrease of position angle from 90$^\circ$ to 50$^\circ$, an opening angle of X-structures increases for about 10$^\circ$. The uniqueness of bulge model is probable due to the fact that such model have large linear sizes in conjunction with high enough opening angle (see Fig.~\ref{fig:models_xz}).}
\label{fig:phi_pos_angle}
\end{center}
\end{figure}

\subsection{Comparison with observations}
\label{sec:sec_observations}
The data on measured flatness or opening angles of X-structures are available for three sample of galaxies \citep{Laurikainen_Salo2016,Ciambur_Graham2016,Savchenko_etal2017}. All three papers exploit quite different methods for analysing the X-structures.  \citet{Laurikainen_Salo2016} prepared unsharp-masked images of galaxies to reveal X-structures. \citet{Ciambur_Graham2016} used Fourier harmonics to describe the deviation of galaxy isophotes from ellipses and demonstrated that the sixth Fourier component traces the X-structure in edge-on galaxies. \citet{Savchenko_etal2017} decomposed the image of an edge-on galaxy into three components, a disc, a bulge and the model of an X-shaped structure. After that, models of the disc and the bulge obtained during the decomposition were subtracted from the original galaxy image. The residual image representing the X-structure was analysed in a manner similar to that proposed for $N$-body snapshots. 
\par
\citet{Laurikainen_Salo2016} list the values characterising the flatness of X-structures which can be transformed into values of opening angles while \citet{Ciambur_Graham2016} and \citet{Savchenko_etal2017} give directly the values of opening angles. The sample by \citet{Laurikainen_Salo2016} has 7 galaxies in common with the sample of \citet{Ciambur_Graham2016} and two common galaxies with the sample by \citet{Savchenko_etal2017}. In all cases opening angles based on unsharp-masked images are on average 10 degrees more than measured by other methods. Particularly curious is the case of the galaxy ESO~443-042, for which analysis of the sixth Fourier harmonics gives an angle of 15$^\circ$ \citep{Ciambur_Graham2016} while measurements based on the unsharp-masked image lead to the angle of 37$^\circ$ \citep{Laurikainen_Salo2016}. Visual inspection of primary and unsharp-masked images suggests that the flatness of X-structures is exaggerated in \citet{Laurikainen_Salo2016}. In the further analysis we excluded the data by \citet{Laurikainen_Salo2016}. 
\par
Fig.~\ref{fig:phi_hist} shows the distribution of 33 galaxies\footnote{Fig.~\ref{fig:phi_hist} shows 36 measurements for 33 galaxies. \citet{Ciambur_Graham2016} give three values of the opening angle for NGC~128 and two ones for NGC~2549 for inner and outer X-structures.} over opening angles obtained as an arctangent of flatness of observed X-structures from samples by \citet{Ciambur_Graham2016} and \citet{Savchenko_etal2017}. A crosshatched region in Fig.~\ref{fig:phi_hist} corresponds to opening angles of our modelled X-structures. Except three individuals, almost all galaxies host X-structures with flatness similar to that in simulated X-structures. The overlapping of two data sets demonstrates a good agreement between observed and modelled X-structures.
Note that the distribution peak of data by~\citet{Savchenko_etal2017} occurs between 25$^\circ$ and 33$^\circ$, which corresponds to thin cool models with $M_\mathrm{h}=1.5-3.0$. Probably, this peak is due to selection effects as in such models X-structures  have a simpler and more symmetrical structure than those in hot or thick enough models. Also note, that observational data demonstrate some small angles (15$^\circ$--20$^\circ$) which are not observed in our simulations. Here we do not want to speculate on any physical reasons behind them as it is possible that they emerge precisely because of different routines of measuring of X-structures. To resolve such inconsistency, a strict comparison of different methods of obtaining X-structure parameters must be done for the same set of observable galaxies or $N$-body models which is beyond the scope of the present work. As for systematic error between different approaches to X-structures mentioned earlier, we cannot obtain any conclusion because the data mostly overlap.
\par 
The obtained numerical trend is that the galaxies possessing massive halos tend to have more flattened X-structures, although we have not reached as small values of opening angles as observed. In addition, the interpretation of observed small angles is complicated because some simulated galaxies demonstrated double X-shaped structures where a secondary X-structure (in sense of the birth time) is more flattened and longer than a short one in the centre of the disc. The present work is probably the first work where the possibility of the formation of such structures in $N$-body models is highlighted. Naturally, it is possible that observed X-structures with small angles are secondary while the first ones just cannot be resolved at the current resolution level. 
\textcolor{black}{
At least for the two observed galaxies, we know that this is not so. \citet{Ciambur_Graham2016} were the first who discovered that NGC~128 and NGC 2549 host multiple (nested) peanuts --- inner and outer ones.
Although these structures are not distinguished by the eye on the galaxy's image, 
\citet{Ciambur_Graham2016} suspected their existence from two explicit peaks on the radial profile of the sixth Fourier harmonic. 
Perhaps such nested peanut-like structures can be associated with the double X-structures that we found in simulations. 
If the line of sight deviates from the minor axis of the bar, both X-structures overlap, and the inner structure will be noticeable in the centre, and the outer structure will appear on the periphery.
\citet{Ciambur_Graham2016} also obtained that outer X-structures, which accompany peanuts, are more flattened and extended which is consistent with the results of our simulations. The inner X-structures in NGC~128 and NGC~2549 have opening angles of 43$^\circ$ and 42$^\circ$, respectively, while for the outer X-structures \citet{Ciambur_Graham2016} obtained 36$^\circ$ (NGC~128) and 34$^\circ$ (NGC~2549).
}
\par
A comparison between observational and simulated data revealed one more feature of X-structures. Observed X-structures do not demonstrate opening angles greater than 50$^\circ$--55$^\circ$. According to~\citet{Savchenko_etal2017}, projection effects (the orientation of a bar with respect to the line of sight) leads to an increase of opening angles by $\sim 10^\circ$. Therefore, if the side-on orientation of a bar gives the upper boundary of opening angles of 40$^\circ$ then, for a rotated bar, observed galaxies must demonstrate even greater opening angles such as mentioned earlier 50$^\circ$--55$^\circ$. Two possible explanations can be applied. On the one hand, Fig.~\ref{fig:models_xz} shows that X-structures with greater opening angles tend to be smaller in linear sizes. If we rotate such an X-structure its rays will drown into the background of high central disc brightness and become indistinguishable from a stellar disc. To test this hypothesis, we rotated some of our models and trace the dependence of opening angles on the bar position angle for end-state snapshots. Fig.~\ref{fig:phi_pos_angle} shows obtained dependencies for some models. For modelled galaxies, X-structure can be traced till the angle of about $20^\circ$ between the bar major axis and the line of sight  but mentioned blurring of X-structure rays is not observed.
On the other hand, the observational sample of 33 galaxies is perhaps not representative and tends to include galaxies with side-on orientation of a bar. Therefore such rather minor inconsistencies between our models and observations can be naturally erased with an increase of sample size. 

\subsection{Comparison with numerical studies}
Such a characteristic as an opening angle is mostly associated with recently obtained observational data and was not given any attention in numerical simulations before. But, of course, the whole picture of vertical bar evolution has been the subject of many different studies (see a review by  \citealp{Athanassoula2016}). In this subsection, we want to discuss what new touches were added by our simulations. 
\par 
First, we obtained interesting patterns of bar plane morphology transformations. It has been recently shown by \cite{Laurikainen_Salo2017} that a barlens/X-shape plane morphology depends on velocity curve steepness. Our simulations of models with a bulge component confirm this result: with a concentrated bulge component, an in-plane bar obtains a barlens morphology (Fig.~\ref{fig:im_bulge}, top row). In addition to that, a direct comparison of end-state of hot and cool models reveals that the same effect can be achieved by an increase of disc thickness or the Toomre parameter (Figs.~\ref{fig:im_thickness}--\ref{fig:im_toomre}, top rows) without a concentrated bulge. Thus, a barlens morphology can be the result of different factors.
\par
Secondly, our simulations extend the conditions under which a recurrent buckling is possible. Such an event is that a bar, which has already established its vertical structure (an X-structure already presents too), starts to lose again the vertical symmetry. It was first described in the work by \cite{MartinezValpuesta_etal2006} based on just one model but a secondary (recurrent) buckling is clearly seen in figure 4 of earlier paper by \citet{Oneil_Dubinski2003}. \citet{Saha_etal2013} considered three different models and in two of them, where a bar had formed, a secondary buckling appeared. In all these works, the duration of the asymmetric state was about $1-2$ Gyrs. From a perspective of our simulations, a potential of a secondary buckling is definitely richer. Hot or thick enough models (the Toomre parameter $Q=2.0$  or flatness $Z_\mathrm{d}/R_\mathrm{d}=0.1;0.2$) buckle for at least $4$ Gyrs. What is more, such a buckling starts relatively late, only from $3-4$ Gyrs. As a consequence, host galaxies are buckled at late stages of evolution, $T_\mathrm{buck} > 8$ Gyr. Naturally, such a long-lasting buckling must manifest itself in observational data. Indeed, some recent observational works \citep{Erwin_Debattista2016,Erwin_Debattista2017,Li_etal2017} have indicated an ongoing buckling by a detailed study of in-plane isophotes asymmetry. The small number of observed buckled galaxies probably indicates that a special combination of initial parameters must take place which is consistent with our simulations. 
\par 
Thirdly, for one model with $Q=1.6$, we obtained that a secondary buckling is even accompanied by the formation of an additional X-structure. \citet{Debattista_etal2017} also give an example of double X-structures in one of their models (their figure 8) and conclude that the main driver of the final morphology is not the thickness of the initial disc, but its in-plane random motion. In our simulations the formation of double X-structures also proceeded in an in-plane hot enough model. Observations confirm that such structures are not an artefact of our simulations and some real galaxies (NGC~128, NGC~2549) have such a complex vertical structure \citep{Ciambur_Graham2016}.
\par
Fourthly, a very noticeable impact on a secondary buckling is caused by a bulge component. In three models with different bulges considered in the present work, secondary buckling is totally suppressed, while for the same model without a bulge it takes place. Prior to this, it has been widely discussed in the literature that the presence of significant gas can also suppress a buckling, at least in simulations \citep{Berentzen_etal1998,Debattista_etal2006}, while \citet{Sotnikova_Rodionov2005} suggest that the presence of a compact, massive, spheroidal bulge could also work. \citet{Sotnikova_Rodionov2005} used rigid spherical component (halo or bulge) in their simulations. Here we prove this result on the base of self-consistent simulations. The same effect can be reached by a very massive dark halo. \citet{Saha_etal2013} give an example of a dark-matter-dominated radially hot
stellar disc. This galaxy model has been evolved for about 12~Gyr during which no buckling event was detected. The variety of conditions, under which gradual, vertically symmetric bar growth is possible, is important for conclusions about frequency of buckled bars among observed galaxies \citep{Erwin_Debattista2016,Erwin_Debattista2017,Li_etal2017}.

\subsection{Secondary buckling}
\label{sec:sec_buckling}
Since a variety of different models considered in the present work is rich enough, we clarified that there are different types of the vertical bar evolution associated with different appearance of buckling. Results described above show that a possibility of secondary buckling and its properties depend on properties of an initial galaxy and more numerical experiments must be done to reveal its abundance. Here we want to enumerate the noticed regularities in the appearance of secondary buckling. Roughly, all models can be divided into four different groups. 
\par

The first family is characterised by a short-lasting buckling at early stages and the calm further evolution without significant changes in the vertical structure. As a rule, this family is associated with a symmetric compact or massive agent, i. e. some subsystem which adds substantial spherical symmetry to a parent galaxy. This role can be played by a bulge or a heavy dark matter halo. A gas component within a bar can also perform this function \citep{Berentzen_etal1998,Debattista_etal2006}. 
\par
The second family which is similar in X-structure appearance to the first family is characterised by two buckling stages. The first buckling is similar to that in the first family. The secondary buckling lasts about $1-2$ Gyr and occurs after $1-2$ Gyr past the first one. Typically for this family, the bar regains its symmetry after the secondary buckling. A calm evolutionary period starts after the end of secondary buckling. In both families, X-structures become visible after the first buckling. After the end of buckling, X-structures tend to lower their opening angles and become larger. As soon as the secondary buckling starts, the averaged over four rays opening angle jumps for about $5^\circ$. After the end of secondary buckling, the X-structure again begins to lower its opening angle.
\par
In general, described two families are associated with thin cool models, $Q=1.2$ and $Z_\mathrm{d}/R_\mathrm{d}=0.05$. In contrary, the third family is associated with thick enough ($Z_\mathrm{d}/R_\mathrm{d}=0.1;0.2$) or hot ($Q=2.0$) models. In such models, the secondary buckling is so vigorous that the bar cannot return to a symmetric state for at least $4$ Gyrs. It is natural to assume that such galaxies, if they exist, must be buckled at our time and must manifest themselves in observations. Probably, the buckling which was detected in real galaxies  \citep{Erwin_Debattista2016,Erwin_Debattista2017,Li_etal2017} is precisely a such type of long-lasting secondary buckling. In general, X-structures have a great opening angle ($35^\circ-40^\circ$) and slightly lower it with time in such galaxies.
\par
As for forth family, it consists only of one member. The model with $Q=1.6$ undergoes a secondary buckling. After that, the model regains its vertical symmetry and \textit{secondary} X-structures appear. It seems that this model demonstrates an intermediate type of secondary buckling that falls between second and third families. Such X-structures manifests themselves in observations too. \citet{Ciambur_Graham2016} found such double X-structures for NGC~128 and NGC~2549.
\par
There exist one more model which does not fit into the general picture. In the model with $M_\mathrm{h}=1$ (the lightest considered halo mass), a bar reformation takes place. A long bar before the reformation and a short bar after the reformation manage to buckle and form an X-structure. If all other models have X-structures which lower their opening angle with time (with an exception of a secondary buckling period), the X-structure in this model has the initial opening angle $36^\circ$ and increases it to about $42^\circ$ during the bar reformation. At all subsequent times, the opening angle stays frozen near this value. Also, a size of the X-structure almost does not change in contrast to the other models.

\textcolor{black}{
The primary buckling event can be understood in terms of bending modes \citep{Raha_etal1991,Merritt_Sellwood1994,Sotnikova_Rodionov2003,MartinezValpuesta_etal2004,MartinezValpuesta_etal2006}. The first buckling distorts the bar's midplane, giving it the saddle shape \citep{Raha_etal1991}, and mostly affects  the central part of a disc. Though the prime driver of the instability is the coupling between the vertical and radial motions this does not explain why buckling happens in central regions. \citet{Sellwood1996} refined the dispersion relation for long-wave perturbations in an inhomogeneous disc by adding an extra term related to the restoring force from the unperturbed disc. This term in the region of rising rotation curve becomes negative and means that the additional expulsive (not restoring) force, which leads to an extra destabilizing effect though the ratio $\sigma_z/\sigma_R$, is above the threshold (e.g. \citealp{Sotnikova_Rodionov2005}).}
\par
\textcolor{black}{
The morphology of the second buckling is quite different from the first one. The buckling hardly affects the midplane but is prominent in the outer bar range and manifests itself as a global vertical asymmetry of the X-structure, which has already been formed. \citet{MartinezValpuesta_etal2006} considered the nature of this buckling in terms of regular 3D orbits bifurcated from the so-called $x1$ family of orbits, which supports the bar (e.g. \citealp{Patsis_etal2002}). By this time, the bar has already grown in the radial direction and in amplitude and as it grows, 3D orbits with high vertical resonances are involved in it. An example of such orbits is shown in figure~8b in the work by \citet{MartinezValpuesta_etal2006}. Asymmetrical orbits are thought to contribute substantially to the vertical asymmetry in the bar during the second buckling. Later, more symmetrical orbits are involved in the bar. Apparently, the global asymmetry in some of our models, especially hot models, is associated with asymmetric orbits, but this requires a separate extensive study.
}
\section{Conclusions}

In the present work, we tried to understand how initial parameters of galaxy models (the number of particles, the mass of a halo or a bulge, the initial thickness and the Toomre parameter) influence the appearance of the vertical bar feature which has an X-shape form seen edge-on. The set of considered parameters was chosen according to its role in the vertical bar evolution which can be described in terms of buckling instability. In total, we consider eighteen different models of galaxies, six of which have the same physical parameter but a different number of particles in a stellar disc and a halo and thirteen have the same number of particles but different values of physical parameters.
\par
To test what impact on X-structures has the vertical relaxation of a stellar disc, we consider a wide range of particle numbers which covers almost two orders of the magnitude, from $N_\mathrm{d} = 2 \cdot 10^5$ to $N_\mathrm{d}=8 \cdot 10^6$. We concluded that, while all discs are subjects to the vertical relaxation, which is well diagnosed by an increase of disc thickness with time, X-structures, at least in their opening angles, nearly coincide within an error of angle measurement and are independent of the number of particles for $N_\mathrm{d,h} > 10^6$. At the same time, an increase in the number of particles is accompanied by a decrease in the measurement error of the opening angle. According to our estimate of the error, to achieve a satisfactory level of accuracy $1^\circ-2^\circ$ comparable to that in observations, the large number of particles, $N_\mathrm{d}=4 \cdot 10^6$, should be used. In accordance with this, we used such a number of particles for all other models with different physical parameters.
\par
By our simulations, we expanded the list of conditions under which the galactic bar undergoes a repeated buckling. In our list there are even models that preserve their buckled structure throughout all simulation time. At the same time, we formulated additional conditions that lead to the gradual symmetrical evolution of the bar in the vertical direction. First of all, this means that the galaxy has a fairly compact bulge with noticeable mass and not only posses substantial mass of gas. Our results are directly related to the search for galaxies with the buckling among the observed galaxies. Until now, the frequency of occurrence of such galaxies was based on a single model the with secondary buckling, which lasted a short time \citep{MartinezValpuesta_etal2006}.
\par
One more by-product but very important result of our simulations is the existence of double vertically symmetric X-structures. Before our work, the appearance of such structures in simulations was revealed only in one study \citep{Debattista_etal2017}. It was noted that the main factor that leads to the emergence of such an unusual final morphology is the in-plane random motion in a disc. Unfortunately, before our work, such structures were not compared with observational data. We note, perhaps for the first time, that this is not an artifact, and it is possible that they were detected in the galaxies NGC~128 and NGC~2549.
\par 
Despite mentioned differences in models evolution, the general picture of dependence of X-structures opening angle on galaxy parameters looks simple enough at least in a qualitative way. An X-structure can obtain high opening angle (30$^\circ-40^\circ$) by many ways including high initial disc thickness, high Toomre parameter, substantial bulge mass, low mass of a dark halo and, finally, by projection effects (a rotation of a bar major axis around the line of sight). In contrary, the list of reasons for X-structures to have low opening angles, $25^\circ-26^\circ$ in our simulations, is substantially poorer. In fact, there are two possible explanations which can be suggested from our simulations. The first one is for galaxy to have a significant halo mass within the optical disc ($M_\mathrm{h}/M_\mathrm{d} \geq 3$). Another one is that the observed X-structures are actually secondary (outer) rays in double X-structures, but in this case the existence of primary (and inner) rays can be checked.
\par
Unfortunately, with the current level of precision of angle measurements and due to differences in observational approaches to X-structures, we can not prove such dependencies by means of observational data. Nevertheless, understanding of general gist of them is still valuable for future researches, and, as we hope, they could be tested with new observational data.

\section*{Acknowledgements}
We thank Research park of St.Petersburg State University <<Computing Center>> where a part of the study was carried out. We thank the anonymous referee for his/her
review and appreciate the comments, which contributed to improving the
quality of the article.


\bibliographystyle{mnras}
\bibliography{article2}



\label{lastpage}
\end{document}